\documentclass[longauth]{aa} 
\usepackage{graphicx}
\usepackage{txfonts}
\usepackage{natbib}
\bibpunct[ ]{(}{)}{;}{a}{}{,}
\def\e20{$\times 10^{20}$}

\def\10K{$\sl{10K}$} 
\def\20K{$\sl{20K}$} 
\def\SG{$\tt{SG}$} 
\def\kmsec{km~s$^{-1}$}
\def\kmsecmpc{km~s$^{-1}~Mpc^{-1}$}

\def\Kpch{$h^{-1}_{70}~Kpc$}
\def\ie{i.e.,~}
\def\vs{versus~}
\def\eg{e.g.,~}

\def\bluefrac{${F_{\rm blue}}$}

\begin{document}
   \title{A journey from the outskirts to the cores of groups I: \\
      Color- and mass-segregation in 20K-zCOSMOS groups \thanks{Based on observations 
      made at the European Southern Observatory (ESO) Very Large 
      Telescope (VLT) under Large Program 175.A-0839}}



\author{
V.~Presotto \inst{1,2}
\and
A.~Iovino\inst{2}
\and
M.~Scodeggio\inst{3}
\and
O.~Cucciati\inst{4}
\and
C.~Knobel\inst{5}
\and
M.~Bolzonella\inst{6}
\and 
P.~Oesch\inst{5,7}
\and 
A.~Finoguenov\inst{8}
\and
M.~Tanaka\inst{9}
\and
K.~Kova\v{c}\inst{5,10}
\and
Y.~Peng\inst{5}
\and
G.~Zamorani\inst{6}
\and
S.~Bardelli\inst{6}
\and
L.~Pozzetti\inst{6}
\and
P.~Kampczyk\inst{5}
\and 
C.~L\'opez-Sanjuan\inst{13}
\and
D.~Vergani\inst{6,24}
\and
E.~Zucca\inst{6}
\and 
L.~A.~M.~Tasca\inst{13}
\and  
C.~M.~Carollo\inst{5}
\and
T.~Contini\inst{11,12}
\and
J.-P.~Kneib\inst{13}
\and
O.~Le~F\`{e}vre\inst{13}
\and
S.~Lilly\inst{5}
\and
V.~Mainieri\inst{14}
\and
A.~Renzini\inst{15}
\and  
A.~Bongiorno\inst{8}
\and
K.~Caputi\inst{16}
\and
S.~de~la~Torre\inst{16}
\and
L.~de~Ravel\inst{16}
\and
P.~Franzetti\inst{3}
\and
B.~Garilli\inst{3,13}
\and
F.~Lamareille\inst{11,12}
\and
J.-F.~Le~Borgne\inst{11,12}
\and
V.~Le~Brun\inst{13}
\and
C.~Maier\inst{5,17}  
\and
M.~Mignoli\inst{6}
\and 
R.~Pell\`o\inst{11}
\and
E.~Perez-Montero\inst{11,12,18}
\and
E.~Ricciardelli\inst{19}
\and
J.~D.~Silverman\inst{9}
\and
L.~Tresse\inst{13}
\and  
L.~Barnes\inst{5}
\and
R.~Bordoloi\inst{5}
\and
A.~Cappi\inst{6}
\and
A.~Cimatti\inst{20}
\and
G.~Coppa\inst{8}
\and 
A.~M.~Koekemoer\inst{21}
\and 
H.~J.~McCracken\inst{22}
\and 
M.~Moresco\inst{20}
\and 
P.~Nair\inst{6}
\and 
N.~Welikala\inst{23}
}

\offprints{Valentina Presotto (valentina.presotto@brera.inaf.it)}


   \institute{
Universit\'a degli Studi dell'Insubria, Via Valleggio 11, 22100 Como, Italy \\ 
 \email{valentina.presotto@brera.inaf.it}
\and
{INAF - Osservatorio Astronomico di Brera, Via Brera, 28, I-20159 Milano, Italy}  
\and
{INAF - IASF Milano, Via Bassini 15, I-20133, Milano, Italy}   
\and
{INAF - Osservatorio Astronomico di Trieste, Via Tiepolo, 11, I-34143 TRIESTE, Italy}  
\and
{Institute of Astronomy, ETH Zurich, CH-8093, Z\"urich, Switzerland}   
\and
{INAF - Osservatorio Astronomico di Bologna, via Ranzani 1, I-40127 Bologna, Italy}  
\and
{UCO/Lick Observatory, Department of Astronomy and Astrophysics, University of California, Santa Cruz, CA 95064}  
\and 
{Max-Planck-Institut f\"ur extraterrestrische Physik, D-84571 Garching b. Muenchen, D-85748, Germany}  
\and 
{IPMU, Institute for the Physics and Mathematics of the Universe, 5-1-5 Kashiwanoha, Kashiwa, 277-8583, Japan}  
\and 
{MPA - Max Planck Institut f\"ur Astrophysik, Karl-Schwarzschild-Str. 1,  85741 Garching, Germany}  
\and 
{Institut de Recherche en Astrophysique et Plan{\'e}tologie, CNRS, 14, avenue Edouard Belin, F-31400 Toulouse, France} 
\and
{IRAP, Universit{\'e} de Toulouse, UPS-OMP, Toulouse, France} 
\and
{Laboratoire d'Astrophysique de Marseille, CNRS-Universit{\'e} d'Aix-Marseille, 38 rue Frederic Joliot Curie, F-13388 Marseille, France}  
\and
{European Southern Observatory, Karl-Schwarzschild-Strasse 2, Garching b. Muenchen, D-85748, Germany}  
\and
{Dipartimento di Astronomia, Universit\`a di Padova, vicolo Osservatorio 3, I-35122 Padova, Italy}  
\and
{SUPA Institute for Astronomy, The University of Edinburgh, Royal Observatory, Edinburgh, EH9 3HJ}  
\and
{University of Vienna, Department of Astronomy, Tuerkenschanzstrasse 17, 1180 Vienna, Austria} 
\and
{Instituto de Astrofis\`ica de Andaluc\`ia, CSIC, Apartado de correos 3004, 18080 Granada, Spain}  
\and
{Instituto de Astrof\`sica de Canarias, V\'ia Lactea s/n, E-38200 La Laguna, Tenerife, Spain} 
\and 
{Dipartimento di Astronomia, Universit\'a di Bologna, via Ranzani 1, I-40127, Bologna, Italy}  
\and 
{Space Telescope Science Institute, 3700 San Martin Drive, Baltimore, MD 21218, USA}   
\and
{Institut d'Astrophysique de Paris, UMR 7095 CNRS, Universit\'e Pierre et Marie Curie, 98bis boulevard Arago, F-75014 Paris, France}  
\and
{Institut d'Astrophysique Spatiale, Batiment 121, CNRS \& Univ. Paris Sud XI, 91405 Orsay Cedex, France}  
\and
{INAF - IASF Bologna, Via P. Gobetti 101, I-40129 Bologna, Italy}  
}

   \date{Received Month Day, Year; accepted Month Day, Year}

 
\abstract
{Studying the evolution of galaxies located within groups may have
important implications for our understanding of the global evolution 
of the galaxy
population as a whole. The fraction of galaxies bound in groups at
z$\sim$0 is as high as 60\% and therefore any mechanism (among the 
many suggested) that could quench star formation when a galaxy enters 
group environment would be an important driver for galaxy evolution.}
{Using the group catalog obtained from zCOSMOS spectroscopic data and
the complementary photometric data from the COSMOS survey, we 
explore segregation effects occurring in groups of galaxies at
intermediate/high redshifts. Our aim is to reveal if, and how
significantly, group environment affects the evolution of infalling 
galaxies.}
{We built two composite groups at intermediate ($0.2 \leq z \leq
0.45$) and high ($0.45 < z \leq 0.8$) redshifts, and we divided the
corresponding composite group galaxies into three samples according to
their distance from the group center. The samples roughly correspond to
galaxies located in a group's inner core, intermediate, and infall
region.  We explored how galaxy stellar masses and colors - working in
narrow bins of stellar masses - vary as a function of the galaxy
distance from the group center.}
{We found that the most massive galaxies in our sample ($\log({\cal
M}_{gal}/{\cal M}_{\odot}) \geq 10.6$) do not display any strong
group-centric dependence of the fractions of 
red/blue objects. For galaxies of lower masses ($9.8
\leq \log({\cal M}_{gal}/{\cal M}_{\odot}) \leq 10.6$) there is a radial
dependence in the changing mix of red and blue galaxies. This
dependence is most evident in poor groups, whereas richer groups do
not display any obvious trend of the blue fraction. Interestingly, mass segregation
shows the opposite behavior: it is visible only in rich groups, while
poorer groups have a a constant mix of galaxy stellar masses as a
function of radius.}
{These findings can be explained in a simple scenario where 
color- and mass-segregation originate from different physical processes. While
dynamical friction is the obvious cause for establishing mass
segregation, both starvation and galaxy-galaxy collisions are plausible 
mechanisms to quench star formation in groups at a faster rate than in 
the field. In poorer groups the environmental effects are caught
in action superimposed to secular galaxy evolution. Their member galaxies 
display increasing blue fractions when moving from the group center to more external regions, 
presumably reflecting the recent accretion history of these groups.}

\keywords{cosmology: observations - galaxies: groups: general - galaxies: evolution}

\authorrunning{Presotto, V. et al}
\titlerunning{Segregation effects in the zCOSMOS-20K group sample}
   \maketitle
%

\section{Introduction}
\label{sect:intro}

The striking bi-modality of the color-magnitude and of the color-mass
diagrams raises important questions about galaxy formation and
evolution. What are the physical processes responsible for the sharp
partition into blue cloud and red sequence galaxies? Does the
environment play a key role in this process by boosting the transition
into the red sequence region?  What are the timescales for this
transition? There is much evidence of correlations between galaxy
properties and their environment, the oldest and best known being the
morphology-density relation \citep[see][, although the first mention of
it dates back to Hubble]{Oemler1974,Dressler1980}. In general, blue,
star-forming, disk-dominated galaxies are located preferentially in
low-density regions, whereas red, inactive, elliptical galaxies favor
high-density regions.

These two distinct galaxy evolutionary families can originate either
from a priori differences set at galaxy formation epoch, the so-called
{\it nature} scenario, or from environmentally driven processes taking
place during the galaxy evolutionary history, the so-called {\it nurture}
scenario. 

The currently accepted $\Lambda$CDM model predicts the hierarchical 
growth of structures: as time proceeds, smaller structures merge to form
progressively larger ones. This process implies that the fraction
of galaxies located in groups progressively increases since $z \sim
1.5$, up to the Local Universe values, where most galaxies are found
in groups \citep{Huchra1982, Eke2004, Berlind2006, Knobel2009}.
As a consequence, at least part of the observed decline of the global
star-formation rate (SFR) from $z\sim1.5$ until today
\citep{Lilly1996, Madau1998, Hopkins2004} could be accelerated by
environmentally driven phenomena. 

In this context group environment plays a dominant
role \citep{Balogh2004, Wilman2005, Iovino2010, Peng2010}, because only a
small fraction of galaxies live in denser environments such as cluster
cores.  There are observational indications that the color
transition from blue to red galaxies proceeds faster in a group than in
the coeval field population, an effect that becomes evident at
redshifts lower than $z \sim 1$ and for galaxies of masses $\log({\cal
M}_{gal}/{\cal M}_{\odot}) \le 10.6$ \citep{Iovino2010, Kovac2010,
Bolzonella2010, Peng2010}.

However, the physical phenomena responsible for accelerating the color
transition from blue to red galaxies within groups are yet to be
described.  While the extreme local densities reached within cluster
cores enable efficient ram pressure stripping of the galaxy cold gas
on timescales of a few Myr \citep{Gunn1972, Abadi1999}, within the groups
different physical processes have been proposed.  On one hand
galaxy-group interactions like 'strangulation', starvation or halo
gas stripping can remove warm and hot gas from a galaxy halo,
efficiently cutting off the star formation gas supply \citep{Larson1980,
Cole2000, Balogh2000, Kawata2008}. Alternatively, mergers/collisions
and close tidal encounters among group member galaxies together with
galaxy-galaxy harassment at the typical velocity dispersion of bound
groups may also result in star-formation quenching \citep{Moore1996}.
These physical processes do not require extreme local densities, and
quench star formation in a more gradual and gentle way on timescales
of several Gyr.

Among the observable effects of these processes are segregation
phenomena, that is, not only differences between group - and field galaxy
properties, but also radial trends of galaxy properties (\eg colors,
morphologies ...) as a function of distance from the group/cluster
center. These phenomena have already been extensively studied in galaxy
clusters, where \eg a strong radial dependence in the star-formation
rate is observed \citep{Hashimoto1999, Balogh1999, Lewis2002,
Balogh2004, Tanaka2004}. The observed quenching of star-formation
activity starts at large cluster-centric distances and low projected
densities, in the so-called infalling regions, and even at large radii
field star-formation values are not yet reached. This result suggests
that galaxy transformation starts to occur in the infalling filaments,
which consist of chains of groups in which field galaxies are affected by
the group environment which changes their star-forming blue
field-like properties into passive, red cluster-like galaxies.
But even if groups seem to be the key environment to search for the
{\it nurture} scenario, still the observational evidence for related
segregation phenomena is quite scarce and holds mainly for the local
Universe \citep{Postman1984, Mahdavi1999, Tran2001, Carlberg2001,
Carlberg2001b, Girardi2003, Dominguez2002, Wilman2009, Bai2010,
Ribeiro2010}.

A complication to consider is the strong correlation between galaxy
properties such as colors, morphologies and star formation, with
galaxy stellar mass \citep{Cowie1996, Gavazzi1996, Blanton2003,
Kauffmann2003, Brinchmann2004, Baldry2004} and the additional correlation
between the galaxy stellar mass itself and environment: galaxies in less
dense environment tend to be less massive than those located in denser
environment \citep{Hogg2003, Kauffmann2004, Blanton2005,
Scodeggio2009, Bolzonella2010}. Thus any study performed on samples of
galaxies containing a wide range of stellar masses cannot distinguish
between true environmental effects and effects simply induced by the
differing mass distributions of galaxies with environment.  To isolate
the true environmental effect, the analysis must be performed in narrow
galaxy stellar mass bins.
Much of the earlier work performed at intermediate/high z was based
on incomplete and/or scarce samples of groups, where often the
statistics was not high enough to perform such an accurate analysis in
small mass bins.

In this paper we will study mass - and color segregation in groups over
wide redshift and galaxy stellar mass ranges using the spectroscopic data
from the recently completed zCOSMOS-Bright, a large survey reaching
out to $z \sim 1$ with a fairly high and uniform sampling rate
\citep{Lilly2007}, and its new group catalog \citep{Knobel2011}. We
will benefit also from the wide range of photometric data available
for the COSMOS survey \citep{Scoville2007, Ilbert2009, Oesch2010}.
Galaxy colors are the easiest parameter to measure among those that
exhibit a distinctive bi-modality, and we selected rest-frame $(U-B)$
color, bracketing the 4000\AA ~break, as a good indicator of the
galaxy average star-formation histories over longer time-scales than
emission line indicators such as \eg [OII].

To shed light on how rapidly and significantly star formation is
suppressed in groups and to overcome the low number
statistics for individual systems (typically 7-8 members per group), we
built stacked groups by co-adding spatial information from group
member galaxies.  This strategy enabled us to establish a statistically
reliable sample and to reveal trends of galaxy properties as a function
of the group-centric distance and of varying group richnesses. 

The paper is organized as follows: in Sect.\ref{sect:Data} we describe
the data of our analysis, including the algorithm chosen to add
the group member galaxies with photometric redshift. 
In Sect.\ref{sect:Mocks} we illustrate the construction of the 
realistic mock catalogs with which we tested our algorithms. 
In Sect.\ref{sect:how_build_sup} we explain how we stacked group member
galaxies to build a composite group. In Sects.\ref{sect:Analysis} and
\ref{sect:Results} we present our analysis and its results, which we
discuss in Sect.\ref{sect:discussion}. Our conclusions are
summarized in Sect.\ref{sect:Conclusions}. A concordance cosmology is
adopted throughout our paper, with $h_{70} = H_0/70$ \kmsecmpc,
$\Omega_{m} = 0.25$ and $\Omega_{\Lambda } = 0.75$.  All magnitudes
are quoted in the AB system throughout.

\section{Data}
\label{sect:Data}

It is widely accepted in the literature that classical
galaxy color/morphology trends in different environments are better
investigated in bins of mass-volume-limited samples \citep{Tasca2009,
Iovino2010, Cucciati2010, Kovac2010, Xue2010, Cooper2010,
Grutzbauch2011}. This strategy enables one to break the degeneracy caused by
the relationships between galaxy stellar masses and
environment and between galaxy stellar masses and colors/morphologies.

The recently completed zCOSMOS-bright survey with its high and
uniform sampling rate offers unique opportunity to explore the
presence/evolution of these trends over a wide range of cosmic time.

\subsection{COSMOS and zCOSMOS surveys}
\label{sect:z_COSMOS}

The COSMOS survey is a large HST-ACS survey, with I-band exposures
down to $\mathrm{I_{AB}}=28$ on a field of 2 deg$^{2}$
\citep{Scoville2007}. The COSMOS field has been the object of 
extensive multiwavelength ground- and space-based observations
spanning the entire spectrum: X-ray, UV, optical/NIR, mid-infrared,
mm/submillimeter and radio, providing fluxes measured over 30 bands
\citep{Hasinger2007, Taniguchi2007, Capak2007, Lilly2007, Sanders2007,
Bertoldi2007, Schinnerer2007, Koekemoer2007, McCracken2010}.

The zCOSMOS survey was planned to provide the crucial high-quality
redshift information to the COSMOS field \citep{Lilly2007}.  It 
benefitted of $\sim 600$ hr of observations at VLT using the VIMOS
spectrograph and it consists of two parts: zCOSMOS-bright, and
zCOSMOS-deep. The zCOSMOS-deep targets $\sim 10000$ galaxies within
the central 1 deg$^2$ of the COSMOS field, selected through color 
criteria to have $1.4 \la z\la 3.0$. The zCOSMOS-bright is purely
magnitude-limited and covers the whole area of 1.7 deg$^2$ of the
COSMOS field. It provides redshifts for $\sim 20000$ galaxies down to
$\mathrm{I_{AB}} \leq 22.5$ as measured from the HST-ACS imaging.  The
success rate in redshift measurements is very high, 95\% in the
redshift range $0.5 < z < 0.8$, and the velocity accuracy is $\sim
100$ \kmsec \citep{Lilly2009}. Each observed object has been assigned
a flag according to the reliability of its measured redshift. Classes
3.x, 4.x redshifts, plus Classes 1.5, 2.4, 2.5, 9.3, and 9.5 are
considered a secure set, with an overall reliability of 99\%
\citep[see][for details]{Lilly2009}.

Our work is based on the the zCOSMOS-bright survey final
release: the so called \20K sample (simply \20K hereafter),
totaling 16623 galaxies with $z\leq 2$ and secure redshifts according
to the above flag classification (18206 objects in total, irrespective
of redshift and including stars).

\begin{figure}
  \centering
  \resizebox{\hsize}{!}{\includegraphics[bb=57 365 358 625]{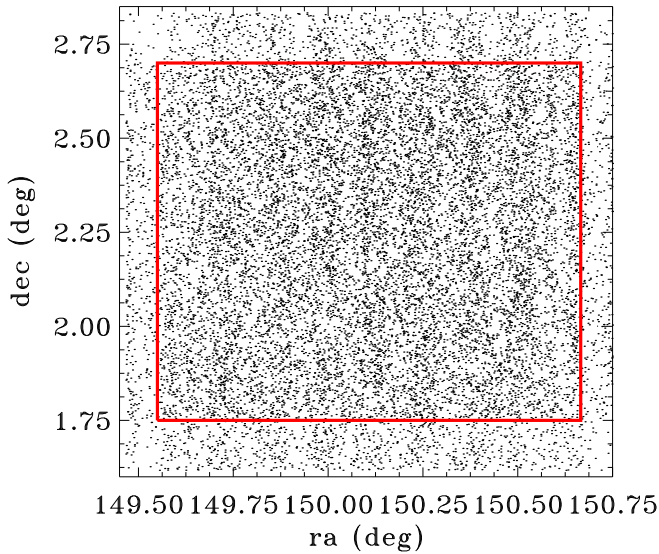}}
\caption{Ra-Dec distribution of the 16623 zCOSMOS-bright galaxies with
secure redshift $z\leq 2$ (the so-called \20K sample). The area within
the red box ($149.55\le ra\le150.67$ and $1.75\le dec\le2.70$) has
a nearly uniform sampling rate of $\sim62\%$.}
\label{fig:20k_zcosmos}
\end{figure}

Fig.\ref{fig:20k_zcosmos} shows the spatial distribution of the \20K
galaxies.  The red square corresponds to the region with the highest
sampling rate, approximately $\sim62\%$ of the parent galaxy
catalog. Its boundaries are $149.55\le ra \le150.67$ and $1.75 \le
dec \le 2.70$.  Within this region are 13619 galaxies with
secure redshift and $z \leq 1$ (15730 objects in total, irrespective
of redshift and including stars) and their sky distribution is
remarkably uniform.

For objects brighter than $\mathrm{I_{AB}}=22.5$ and without secure
spectroscopic redshift, the wealth of ancillary photometric data
provided by the COSMOS survey provides good quality photometric
redshifts \citep{Ilbert2009}.  Based on a comparison with the zCOSMOS
spectroscopic redshifts, \citet{Ilbert2009} estimate an accuracy of
$\sigma_{\mathrm{zphot}}=0.007\times(1+z_{\rm s})$ for galaxies
brighter than $\mathrm{I_{AB}}=22.5$.  
Applying the ZEBRA code \citep{Feldmann2006} to 30 bands,
\citet{Oesch2011} obtains a similar accuracy (private communication).

In our analysis we used photometric redshift values obtained by
the ZEBRA code.

For all galaxies brighter than $\mathrm{I_{AB}}=22.5$, absolute
rest-frame magnitudes and stellar masses were obtained using standard
multi-color spectral energy distribution (SED) fitting techniques,
using the secure spectroscopic redshift, if available, or the
photometric one. Rest-frame absolute magnitudes were obtained using
the ZEBRA code \citep[see][for the details of the
code]{Feldmann2006}, while stellar masses were obtained using the
hyperzmass code \citep{Pozzetti2010, Bolzonella2010}.From the
available stellar population synthesis libraries we adopted those of
\citet{Bruzual2003}, assuming a Chabrier initial mass function
\citep{Chabrier2003}.

\subsection{The spectroscopic group catalog}
\label{sect:group_cat}

The group catalog used in this paper is a subset of the \20K group
catalog described in \citet{Knobel2011}, \citet[see also][for an
earlier version of the catalog]{Knobel2009}.  The \20K group catalog
consists of 1496 groups with at least two spectroscopic member galaxies
(188 with at least five spectroscopic members). \citet{Knobel2011} uses a
``multi-pass procedure'' to achieve an impressive quality in group
reconstruction, as tested using realistic mock catalogs. This method,
when combined with the standard friends-of-friends (FOF) algorithm,
yields for the resulting group catalog values of {\it completeness}
(\ie fraction of real detected groups) and {\it purity} (\ie fraction
of non-spurious groups) that are extremely good and stable as a
function of both redshift and number of members observed in the
reconstructed groups.  Typical values of these two quantities, for
groups reconstructed with five or more spectroscopic observed members,
are around $\sim 80$\% at all redshifts and do not decrease
substantially for groups with lower number of observed members. The
interloper fraction, \ie the fraction of field galaxies erroneously
classed as group members, always remains below $\sim 20$\% at all
redshifts for groups reconstructed with more than five spectroscopic
observed members, with only a slight increase for groups with lower
number of observed members. Another point worth noticing is that the
algorithm to detect groups treats each galaxy as a point in
Ra-Dec-redshift space, therefore avoiding any
interloper/completeness dependence on galaxy properties such as colors
or masses \citep[see][for more details]{Knobel2011}.

In this paper the analysis is restricted to groups with at least five
spectroscopically observed members that are located within the high
sampling rate box introduced in Fig.\ref{fig:20k_zcosmos}.  From now
on we will call this sample the spectroscopic group sample: it
totals 178 groups and 1437 group member galaxies at $z \leq 1$.  Our
choice enables us to work with groups that have best values for purity
and interloper fraction, and to secure a reliable definition of group
center and radius. These two parameters are crucial to build the composite 
group and for our algorithm which retrieves group members without 
spectroscopic redshift informationd (see later Sect. \ref{sect:add_photo}
and \ref{sect:how_build_sup}).

\subsection{The spectroscopic field sample}
\label{sect:field_sample}

To define the field galaxy sample, we started by selecting \20K
galaxies located within the high sampling rate box and outside {\it
any} of the reconstructed groups of \citet{Knobel2011}.  We therefore
discarded from this sample galaxies located in pairs, triplets and
quadruplets, \ie members of the groups with fewer observed members are not
considered in our science analysis.

To perform the fairest comparison between group and field samples, we
took into account the possibility of spurious trend introduced by
residual group contamination or by the different redshift ranges
covered by the group/field galaxy samples. Galaxies lying in the
closest proximity of groups could be contaminated by spectroscopic
group members missed by the group-finding algorithm.  In addition, the
redshift distribution of the spectroscopic group catalog is far from
being uniform, displaying prominent peaks, especially at low redshift
where the \20K field of view limits the cosmic volume explored and we
need to consider the appropriate coeval field population. 

To take into account these two factors, we further more restricted the field
sample to galaxies located within velocity distances $2000 \leq \vert
\Delta v \vert \leq 5000$ km s$^{-1}$ from the spectroscopic group sample,
- therefore following the same redshift ditribution of group
member galaxies - and with radial projected distances $R > 4 \times
R_{fudge}$ from any group of \20K group catalog, $R_{fudge}$ being an
estimate of the virial radius provided by \citet{Knobel2011} (see
Sect.  \ref{sect:super_group} for details).  From now on we will
call this set of galaxies the {\it field sample}, totalling 6556
galaxies at $z \leq 1$.

We also introduced a complementary set of field galaxies that we
call {\it near-field} galaxies. These are \20K galaxies within
the high sampling rate box that do not belong to {\it any} of the
reconstructed groups of \citet{Knobel2011}, but with velocity
distances $\vert \Delta v \vert \leq 2000$ km s$^{-1}$ and radial
projected distances $R \leq 4\times R_{fudge}$ from at least one group
of the spectroscopic group sample. The {\it near-field} sample so
defined totals 1694 galaxies at $z \leq 1$ and contains, by
definition, galaxies located in the close proximity of the
spectroscopic group sample.  In Sect.\ref{sect:near_far_field} we will
use this sample to check for possible environmental effects extending
outside group radii, \eg color differences of {\it near-field} galaxy
population with respect to the general field sample.

\subsection{Adding photo-zs: the spec+photo-z group catalog}
\label{sect:add_photo}

For each group the number of available member galaxies down to
$\mathrm{I_{AB}} = 22.5$ is limited by the incomplete sampling rate of
\20K. To increase this number, we took advantage of the
exquisite quality of the photometric redshifts available in the COSMOS
field, see Sect.\ref{sect:Data} to incorporate in our analysis
photometric redshifts for galaxies brighter than $\mathrm{I_{AB}} =
22.5$ and without reliable spectroscopic data.  A higher number of
group member galaxies enables one to improve centering
and richness estimates for each group, quantities crucial to properly center and
rescale distinct groups to build a composite one \citep[see
\eg][]{Carlberg1997}.

In \citet{Knobel2011} a probability approach was adopted to retrieve
member galaxies brighter than $\mathrm{I_{AB}}=22.5$ and that have no
spectroscopic information.  To each galaxy a probability, $p_{in}$, of being
part of a group was assigned, depending on its projected radial and
velocity distance from the group center (we refer the reader to the
paper by \citet{Knobel2011} for a more detailed description of the
adopted method). The drawback of this approach is that each galaxy may
have multiple associations to different groups.

To overcome this drawback, we developed a slightly different strategy,
whose main advantage is that it assigns each galaxy only
one spectroscopic group, thus avoiding multiple assignments of a
galaxy to different groups, and the need to adopt an arbitrary
probability cut-off to bypass this problem.

For a detailed description of our algorithm we refer the reader to
Appendix \ref{app:algorithm}, while in Sect. \ref{sect:Mocks} we will
present extensive tests that we performed on mock catalogs to check the
reliability of the final spec+photo-z group catalog.

Suffice is to say that we chose the selection function to
identify putative photometric-redshift members in a way to not only
keep the fraction of interlopers as low as possible, but also to
avoid introducing any radial dependency of the interloper fraction.
The last point is important because we will be looking for radial
dependencies of galaxy properties.

As already mentioned in Sect.\ref{sect:group_cat}, we chose a
conservative definition of the spectroscopic group sample, restricting
ourselves to only 178 groups detected with at least five spectroscopic
members within the the high sampling rate box introduced in
Fig.\ref{fig:20k_zcosmos} box ($149.55\le ra \le150.666$ and $1.75 \le
dec \le 2.7$). Within this area and up to $z = 1.0$ there are 13619
galaxies with reliable spectroscopic redshift and 11994 with an
estimated photometric redshift.

Our algorithm adds another 684 member galaxies with photometric
redshifts to the already existing 1437 spectroscopic group member
galaxies, and from now on this is the group sample we will
use. The final number of groups with more than 10(15) members after
applying our algorithm is twice(three times) that in the
spectroscopic group catalog, \ie there are 78(41) groups instead of
the original 39(14) groups. The number of groups with more than 20
members is six times the original one: 25 groups instead of the
original four groups.

As a final point we notice that we repeated all analyses presented
in this paper considering only galaxies from the spectroscopic group
catalog and our results remained entirely unchanged, albeit at a lower
significance.  

\section{The zCOSMOS mock catalogs}
\label{sect:Mocks}

The use of realistic mock galaxy catalogs is important for assessing the
reliability of the algorithm we adopted to produce the spec+photo-z
group catalog and to validate the procedures we chose to define group
centers and richnesses (see Sect.\ref{sect:add_photo} and
Sect.\ref{sect:how_build_sup}).

We took advantage of the 24 COSMOS mock light-cones provided by
\citet{Kitzbichler2007}. These mock light-cones are based on the
Millennium DM N-body simulations of \citet{Springel2005} and use
semianalytic recipes of \citet{Croton2006} as updated by
\citet{DeLucia2007} for populating the simulations volume with
galaxies.

From each of these 24 light-cones we extracted three different types
of mock catalogs:

\begin{enumerate}
 \item The 40K mock catalogs: 100\% complete to
 $\mathrm{I_{AB}}=22.5$. In these catalogs all galaxies brighter than
 $\mathrm{I_{AB}}=22.5$ are spectroscopically observed with a 100\%
 success rate. We added to each galaxy redshift an error of 100 \kmsec
 to account for the typical zCOSMOS spectroscopic redshift error as
 estimated from observations \citep[see][]{Lilly2009}.

 \item The 20K mock catalogs: mimicking the 20K zCOSMOS spectroscopic
 sample. We applied the same observational strategy adopted to select
 the spectroscopic zCOSMOS targets: using the slit
 positioning algorithm SPOC on the 40K catalogs, see \citet{Bottini2005}, and accounting
 for the spectroscopic redshift failures by including the same
 redshift success rate as the real data.

 \item The 20K+photo-zs mock catalogs: mimicking the data set we used
 in our analysis. The spectroscopic galaxies are those listed in the
 20K mock catalogs, while a photometric redshift is provided for the
 remaining galaxies of the 40K mock catalogs.  For the photometric
 redshift galaxy sample we reproduced the photometric redshift error
 $\sigma_{\mathrm{zphot}}=0.007\times(1+z_{\rm s})$.  We also took
 into account the presence of catastrophic failures in estimating
 photo-z, that is, the excess of galaxies with errors larger than
 $\Delta$phot-z $= 3\sigma_{\mathrm{zphot}}$ with respect to the
 simple Gaussian distribution. For the set of photometric redshift 
 adopted in our analysis we were able to estimate a percentage of $\sim$
 8\% (by using the \20K subset flagged 4.x or 3.x and comparing their
 spectroscopic redshift to their photometric redshift).  Group members
 with such high values of phot-z error cannot be retrieved by our
 algorithm and are a considerable source of incompleteness in group
 reconstruction. We chose a fairly conservative approach and also
 considered catastrophic errors of 10\% in the 20K+photo-zs mock catalogs, 
 by randomly permuting the photometric redshifts for 10\% of the galaxies
 while keeping the galaxy ra-dec fixed.
 coordinates.
\end{enumerate}

We applied to the 20K mock catalogs the same group finding algorithm
used for the \20K sample \citep[see][]{Knobel2009}. We then selected
groups with at least five spectroscopic members located within the
high sampling rate box introduced in Fig.\ref{fig:20k_zcosmos} and
applied to the 20K+photo-zs mock catalogs the algorithm described in
Sect.\ref{sect:add_photo}.

\begin{figure*} 
 \centering
 \includegraphics{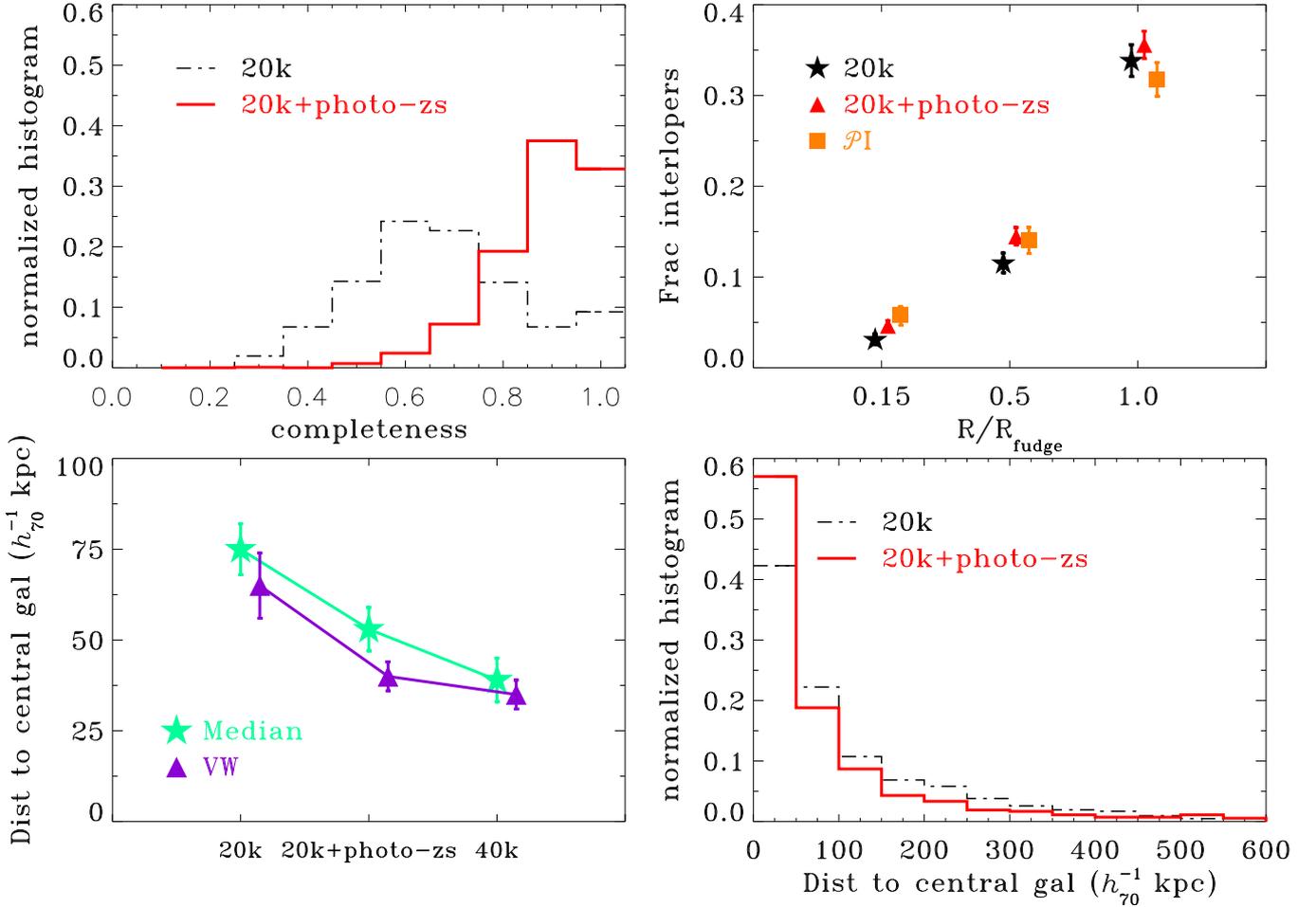}
 \caption{\footnotesize{Summary of the results obtained with our
 algorithm.  {\it Top left}: completeness distribution (see
 text for definition) for the 20K mocks (black dot-dashed line) and
 for the 20K+photo-zs mock groups (red solid line). {\it Top right}:
 fraction of interlopers as a function of normalized group-centric
 distance for the 20K/20K+photo-zs mock groups (black stars and red
 triangles, respectively). Orange squares refer to the fraction of
 interlopers, ${\cal PI}$, for real data, as calibrated on the mocks,
 see text for details. {\it Bottom left}: for different mocks as
 indicated on x-axis, the median distance to the central galaxy of the
 VW center (violet triangles) and the median center (cyan stars). {\it
 Bottom right}: Distribution of the distance of the VW center to the
 central galaxy position for the 20K mocks (black dot-dashed line) and
 for the 20K+photo-zs mocks (red solid line).}}
 \label{fig:results_alg_mock}
\end{figure*}

The COSMOS mock light-cones provide dark matter halos IDs that can
easily be used to identify {\it real} groups and {\it real} group
members, \ie the set of galaxies located within the same dark matter
halo in each mock catalog \citep[see also][]{Knobel2009}.  If we
define completeness as the ratio of the reconstructed group members in
the 20K/20K+photo-zs mock catalogs to the total number of real group
members in the 40K mock catalogs, the improvement introduced by our
algorithm is shown in the top left panel of
Fig.\ref{fig:results_alg_mock}. This panel shows the distribution of
the completeness for all groups of the 20K mocks and for those
obtained after applying our algorithm to the 20K+photo-zs mocks (black
dot-dashed line and red-solid line, respectively).  We were able to
improve the median completeness from 67\% of the 20K mock catalog up
to 90\% in the 20K+photo-zs mocks: the number of groups that are 100\% complete
is three times larger than using only the 20K mocks. As a consequence,
the group richness, defined as the number of members brighter
than an adopted rest-frame absolute magnitude cut-off, is also easier to
recover in a reliable way. For more than half of the cases the
richness as measured for reconstructed groups in the 20K+photo-zs
mocks equals the same quantity as obtained from the 40K mock catalogs.

Our algorithm achieves this remarkable result while adding a
negligible fraction of interloper members, \ie galaxies that do not share
the same dark matter halo in the 40K mock catalogs. For half of the
groups we added less than 3\% of new interlopers with respect to the
total members in the 20K+photo-zs catalog, therefore attaining the
same interlopers fraction as in the spectroscopic group catalog.

We also checked for any dependence of the interloper fraction on the
normalized group-centric distance
$\mathrm{R_{gal}}/\mathrm{R_{fudge}}$, $R_{fudge}$ being an estimate
of the virial radius provided by 
\citet[][see Sect. \ref{sect:super_group} for details]{Knobel2011}. 
For this test we divided each
group into a central part and two concentric intermediate and external
rings, as in Sect.\ref{sect:super_group}, and adopted exactly
the same mass and redshift limits adopted subsequently in our analysis, see
Sect.\ref{sect:Analysis}.

In the top right panel of Fig.\ref{fig:results_alg_mock} we show the
fraction of interlopers in each of these three regions for the
20K/20K+photo-zs mock groups (black stars and red triangles,
respectively), as obtained using the low-z, mass-limited, mock group
samples.  Notice that the trend introduced by the group-finding
algorithm in the 20K mock reconstructed groups is not modified by
adding photo-zs members.  In the same panel the orange squares
display the fraction of interlopers in the \20K sample
spectroscopic group catalog, estimated using the probabilities,
$p_{in, i}$, associated to each observed group spectroscopic member:
 
\begin{equation}
{\mathcal PI} = 1-\sum_{i=1}^{N_{tot,obs}}{p_{in, i}/N_{tot,obs}},
\label{eq:f_int_all}
\end{equation}

as provided by \citet{Knobel2011}.  These values have been calibrated
in \citet{Knobel2011} using simulations, and therefore are by
construction agree well with the interloper fraction
estimated from 20K mock groups. In turn, because adding phot-z members does
not alter the trends significantly, these values agree well
with the interloper fraction for the 20K+photo-zs mock groups.  The
picture does not change when plotting the same quantities for the
high-z, mass-limited, mock group samples, or when selecting subsets
of groups according \eg to their richness.  Therefore we always
used the ${\cal PI}$ values obtained from spectroscopic group members to
estimate the interlopers' contamination as a function of the distance
from the group center for our spec+photo-z group catalog, see
Sect.\ref{sect:color_seg}.

\section{Building the stacked group}
\label{sect:how_build_sup}

To explore galaxy properties as a function of the distance from the
center of the group, we needed to build ensemble systems, because the scarcity
of individual group members prevents a detailed analysis of each
group.  In this section we illustrate in detail the steps of building the
so-called {\sl stacked-group}: a composite group obtained by spatially 
co-adding all group member galaxies (simply \SG\, from now on).

The two main ingredients to build a \SG\, are precise re-centering and
scaling of all available groups. It is therefore extremely important
that each group center and richness is defined as reliably as
possible, so that the trends we are searching for are not smoothed
out. We will discuss precise definition of both quantities in this
section. We remind the reader that from now on any number quoted,
unless explicitly stated, includes both spectroscopic and photometric
redshift group member galaxies, as obtained from the algorithm
discussed in Sect.\ref{sect:add_photo} and which we describe in more 
detail in Appendix
\ref{app:algorithm}.

\subsection{Group centering} 
\label{sect:center}

A good group center definition is essential to our science analysis,
because we will be searching for radial trends that can be easily erased by
errors in group centering.  After adding photo-zs, as discussed in the
previous section, 50\% of the groups in our sample possess more than
nine members which makes group center definition more robust. However
the simple methods of estimating group centers, such as the median of
members coordinates, provide only rough estimates of the group center,
especially for the numerically poorer groups.  We therefore tried an
alternative strategy, taking into account sky-projected group galaxy
densities.

Using the 2D-Voronoi areas as proxy for density measurement, we
defined the Voronoi-weighted center (VW center from now on) as

\begin{equation}
	\mathrm{ra_{VW}}=\frac{\sum^N_{i=1}\mathrm{ra_i} /
	\mathrm{A_{V,i}}}{\sum^N_{i=1}1/\mathrm{A_{V,i}}}, \quad
	\mathrm{dec_{VW}}=\frac{\sum^N_{i=1}\mathrm{dec_i} /
	\mathrm{A_{V,i}}}{\sum^N_{i=1}1/\mathrm{A_{V,i}}}.
\label{eq:VW_center}
\end{equation}

where $\mathrm{A_{V,i}}$ is the 2D-Voronoi area associated to the i-th
galaxy member, that is, the projected area containing all points closer
to the i-th galaxy than to any other member galaxy. We used galaxies
located outside $3\times\mathrm{R_{gr}}$ (where $\mathrm{R_{gr}}$ is
the radius of the minumum circle containing all group members) and
within $1\times\mathrm{\sigma_{zphot}}$ to avoid divergence of
2D-Voronoi areas for galaxies located at the periphery of groups.

This way galaxies that are located in group denser regions will have a
smaller $\mathrm{A_{V,i}}$ and they will weigh more, while those that
are in less dense regions will affect the center
determination less. This method thus provides a center for the group, which
is located by definition in the area of greatest galaxy over-density,
and is not affected by the details of the spatial distribution of
galaxies at the outskirt. For a similar approach see \citet{Diaz2005}.

We used our set of mock catalogs to test the advantages of this center
definition with respect to simpler ones, like the median of the member
galaxies coordinates (median center from now on). We assumed the position 
of its central galaxy as fiducial center for each group, as
provided by the mocks.

In the bottom left panel of Fig.\ref{fig:results_alg_mock} we show the
median distance of the VW center (violet triangles) and that of the
median center (cyan stars) for each of the three mock catalogs defined
in Sect.\ref{sect:add_photo}. Error bars show the rms among mock
catalogs extracted from the different 24 light-cones. We note that the
VW center provides a better estimate of the center with respect to the
median center on average.  Furthermore, the VW centers, when applied to
the groups whith photo-zs added using our our algorithm, are nearly
indistinguishable from those obtained when all members down to
$\mathrm{I_{AB}=22.5}$ possess spectroscopic redshift. The median
value of the distance of the VW centers from the group central galaxy
is 40 \Kpch\, for 20K+photo-zs mocks, with an improvement of nearly
40\% in centering with respect to the 20K mocks.

In the bottom right panel of Fig.\ref{fig:results_alg_mock} we show
the distance histogram of the VW center to the central galaxy
position for the 20K mocks as a black dot-dashed line and for the
20K+photoz mocks as a red solid line. The improvement in group centering
obtained when adding photo-z members is quite obvious.

We therefore adopted the VW method to define the center of each group.

\subsection{Group rescaling}
\label{sect:super_group}  
 
The procedure of stacking groups of different sizes and masses into an
ensemble system requires rescaling of individual galaxy
group-centric-distances.  In studies of galaxy clusters, projected
cluster-centric-distances R are generally rescaled with $R_{{\rm
vir}}$ or $R_{{\rm 200}}$, whose estimate is proportional to cluster
velocity dispersion $\sigma _{{\rm v}}$, that is, a proxy of cluster
mass \citep{Carlberg1997,Biviano2002,Katgert2004}. However, the problem
is not trivial when dealing with galaxy groups, where the
uncertainties in the estimate of velocity dispersions, masses, size,
and the group dynamical state in general are larger, because of the small
number of group members.

In the literature there are different approaches in rescaling R for
groups, using 1) the virial radius $R_{{\rm vir}}$ or $R_{{\rm 200}}$,
2) an estimate of the rms of the position of member galaxies $R_{\rm
{H}}$, and 3) sometimes radial distances are not rescaled at all
\citep[see][for a detailed review]{Girardi2003}.

Our groups span a wide range of sizes, and therefore a rescaling of
physical distances seemed unavoidable.  We decided to use $\mathrm{R_{fudge}}$ 
as the scaling factor, provided by \citet{Knobel2011}.
This {\it fudge} quantity, as many other ones correlating with the
observed group richness, was estimated and calibrated using our
realistic mock catalogs. In brief, given an observed group at redshift
$z$ with richness $\cal N$, defined as the number of members brighter
than an adopted rest-frame absolute magnitude cut-off, its
$\mathrm{R_{fudge}}$ corresponds to the mean $R_{{\rm vir}}$ among all
reconstructed mock groups wjth the same $\cal N$ and redshift
\citep[see][for more details on how this quantity is
calculated]{Knobel2011}.  The quantity $\mathrm{R_{fudge}}$
correlates with $M_{halo fudge}$, an estimate of the mass of the
group well, which additionally shows its relevance for our analysis
\citep[see][for more details on how both these quantities are
estimated]{Knobel2009, Knobel2011}.

Because our goal is to distinguish property of galaxies located in
regions with different physical properties rescaling by
$\mathrm{R_{fudge}}$, a quantity related to $R_{vir}$, suits our
needs well.  Indeed, the virial radius is a scaling factor for many
timescales of different processes such as the crossing time, the
relaxation time or the merging time
\citep{Boselli2006,Weinmann2006}. All galaxies that are inside the
virial radius are experiencing the group potential effects either for
the first time or many times. In contrast, those galaxies that are
outside the virial radius are a mixed population of both in-falling
galaxies and galaxies that once passed through the virial radius but
now are in the outskirts, the so-called back-splash population
\citep{Gill2005}.

Before stacking groups, we therefore rescaled each member galaxy
distance to the VW center, $\mathrm{R_{gal}}$, with the corresponding
$\mathrm{R_{fudge}}$ of its group.  Below we will use only scaled
distances, ${\cal R}$, unless otherwise specified.

We add a final caveat: when discussing our results, we should
take into account projection effects.  We observed the 2-D projection
of a 3-D distribution of member galaxies. Assuming a spherical
distribution, this implies that \eg the inner observed region includes
galaxies located in the outer group shells that are located along the line
of sight of the group central part. Hence a fraction of galaxies
observed, in projection, in the inner region actually belongs to the
outskirts. These projection effects will tend to smooth the radial
trends we are looking for, so that any observed trend is a lower limit
for the real trend present in 3-D.  Vice versa in the external regions
of groups the contamination by field galaxies, on average bluer and
less massive than group galaxies, will tend to introduce spurious
segregation trends, and we need to account for them carefully.

\begin{figure*}
    \centering
  \resizebox{\hsize}{!}{\includegraphics[bb=56 423 542 584]{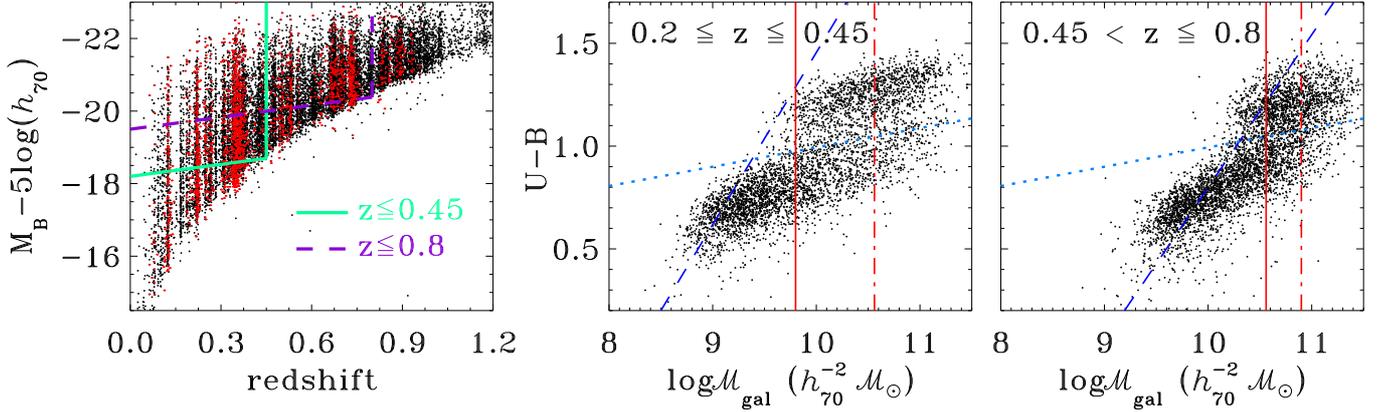}}
\caption{{\it Left panel}: redshift distribution of the zCOSMOS-bright
galaxies (black). Red points represent group member galaxies with
spectroscopic and photometric redshifts as obtained from our
algorithm. The cyan solid line and the violet dashed line correspond to the
two different magnitude cut-offs adopted to define the volume-limited 
samples of the low- and high-redshift bin respectively, see text for details.  
{\it Central and right panel}: $(U-B)$ rest-frame color versus mass for the lowest and the
highest redshift bin. The blue dashed line corresponds to
the color-dependent ${\cal M}_{\rm cut-off}$, while the red solid line
corresponds to the fixed ${\cal M}_{\rm cut-off}$ for our mass
volume-limited sample. The red dot-dashed lines highlight the mass
ranges adopted in the mass-segregation analysis, see
Sect.\ref{sect:mass_seg} for details.  The cyan dotted line
corresponds to the separation between red and blue galaxies 
(see text for its precise definition).}
\label{fig:Mag_z_plus_UB_mass_dist}
\end{figure*}

\section{Analysis}
\label{sect:Analysis}

\subsection{Selecting mass volume-limited samples}
\label{sect:vol_mass_lim_sample}

We focused our analysis on two redshift bins: $0.2\leq z \leq 0.45$
and $0.45 < z \leq 0.8$, where we defined the classical volume-limited
samples taking into account the luminosity evolution of individual
galaxies. Following \citet{Zucca2009}, we adopted a linear evolution
with redshift: $M^*_{B \rm ev} = -20.3 - 5~\log~h_{70}-1.1~z$ to
parametrize the evolution of $M^*_{B}$ of the luminosity function. The
corresponding evolving cut-off magnitudes are $M_{\rm cut-off} =
M^*_{B {\rm ev}} + 2.1(+0.8) $ for the low(high) redshift bin.  
For $0.2\le z \le 0.45$ the volume-limited sample
consists of 829 out of 1128 total galaxies, belonging to 79
groups. For $0.45 < z \le 0.8$ it consists of 510 out of 660 total
galaxies, belonging to 64 groups (see Tab.\ref{tab:N_gals}).  The
total volume-limited field sample consists of 1869(2893) galaxies,
while the {\it near-field} volume limited sample consists of 683(612)
galaxies for the low(high) redshift bin.  In the left
panel of Fig.\ref{fig:Mag_z_plus_UB_mass_dist} we show the $M^*_{B}$
\vs redshift distribution of the total galaxy sample (black points)
and that of both spectroscopic and photometric redshift group member
galaxies (red points). The cyan solid line and the violet dashed line
correspond to the magnitude cut-offs defining the low- and 
high-redshift volume-limited samples. In the following, the group richness
$\cal N$ for each group is defined as the number of (phot+spec-z)
member galaxies surviving to the more conservative absolute rest-frame
magnitude cut-off: $M^*_{B{\rm ev}} +0.8$, unless explicitely stated. 
This quantity correlates, albeit with a large scatter, with the mass
of the halo where the group resides and therefore is a good proxy for
it \citep[see][]{Knobel2009}.

The flux-limited target definition of zCOSMOS-bright, $\mathrm{I_{AB}}
\leq 22.5$, translates into a B-band rest-frame selection at $z \sim
0.8$. Therefore the \20K\, galaxy sample, when rest-frame B-band
selection is adopted, is free from significant color-dependent
incompleteness in $(U-B)$ rest-frame colors up to $z \sim 1$.  However
the $(U-B)$ rest-frame color completeness in the B-band rest frame selection
does not imply completeness in mass selection: the B-band rest-frame
selection is biased toward blue, low-mass galaxies, while missing the
corresponding red, equally low-mass ones.  Environmental trends
observed in samples selected using rest-frame B-band magnitudes could
therefore be simply the results of this incompleteness coupled with
different galaxy mass distributions in different environments
\citep{Bolzonella2010}.

To separate true environmental effects from mass-driven ones, we
used in our analysis mass volume-limited samples, that is, samples
complete down to a fixed galaxy mass cut-off. To obtain them, we
followed the same approach as in \citet{Iovino2010}. In brief, we first
calculated the limiting stellar mass for each galaxy {in the \20K
sample}, \ie the stellar mass it would have at its spectroscopic
redshift, if its apparent magnitude were equal to the limiting
magnitude of our survey: $\rm \log({\cal M}_{lim}(z_{gal}))=\log({\cal
M}_{gal}) + 0.4\, (\mathrm{I_{AB}}-22.5)$. We then used these
estimated limiting masses to define, in bins of $(U-B)$ rest-frame
colors for each redshift bin, the mass ${\cal M}_{\rm cut-off}$ below
which 85\% of galaxies of that color lie. We fitted ${\cal M}_{\rm
cut-off}$ to obtain a color-dependent mass limit cut-off.  The value
of ${\cal M}_{\rm cut-off}$ for the reddest galaxies in each redshift
bin is the one that we used as the limiting mass for that bin.

In the central and right panel of
Fig.\ref{fig:Mag_z_plus_UB_mass_dist} we show the $(U-B)$ rest-frame
color versus the stellar mass for the lowest and highest redshift bin
respectively. The blue dashed line shows the color-dependent ${\cal
M}_{\rm cut-off}$, while the red solid line shows the value chosen to
define mass-limited samples: 
$\log({\cal M}_{gal}/{\cal M}_{\odot}) \ge {\cal M}_{\rm cut-off}=9.8$ and 
$\log({\cal M}_{gal}/{\cal M}_{\odot}) \ge {\cal M}_{\rm cut-off}=10.56$ 
for the lowest and highest redshift bins. 

To define the mass-dependent color cut separating the blue and 
red galaxies, we performed a robust fit of the red sequence as 
a function of the galaxy stellar mass in the high-z bin, where
a large number of observed galaxies displays a prominent and
well defined red sequence. The color cut was then obtain by shifting 
the fitting line by 
$2\cdot rms_{red}$, where $rms_{red}\sim0.08$ is the dispersion 
of the red galaxies along the red sequence. We adopted the same 
color cut for the low-z bin.
Numerically, the stellar mass dependent color cut is

\begin{equation}
(U-B)=0.094\cdot \log({\cal M}_{gal}/{\cal M}_{\odot}) + 0.05,
\label{eq:color_cut}
\end{equation}

and it is shown by the cyan dotted lines in
Fig.\ref{fig:Mag_z_plus_UB_mass_dist}.

We tested that our results do not change if we apply a constant
color cut, $(U-B)$ = 1, to separate red and blue galaxies, a simpler
definition that corresponds equally well to the dip of the bimodal 
distribution.

For the lowest redshift bin the final group mass-complete sample
contains 571 galaxies, while for the highest redshift bin it contains
265 galaxies. The mass-complete field samples consist of 743(728)
galaxies for the lowest(highest) redshift bin, while the
{\it near-field} samples consist of 293(211) galaxies for the
lowest(highest) redshift bin.

\begin{table} 
\caption{Number of volume-limited and mass-volume-limited
(spec+phot-z) group member galaxies. In brackets we report the number
of spectroscopic-only group members. The number of groups containing
these galaxies is listed in the last column.}
\label{tab:N_gals}
\centering
\begin{tabular}{l c c c}
\hline
 Redshift &   Vol-lim & Vol-Mass-lim &     \\
    & $N_{gals}$  & $N_{gals}$  & $N_{gr}$    \\
\hline
 $0.2 \leq z \leq 0.45$ &   829 (570) & 571 (410) & 79\\
 $0.45 < z \leq 0.8$ &  510 (391) & 265 (200) &  64 \\   
\hline
\end{tabular}
\end{table}

\subsection{Low-z and high-z stacked-groups}
\label{sect:two-super}

\begin{figure*}
 \centering
\resizebox{\hsize}{!}{\includegraphics[bb=63 433 545 598]{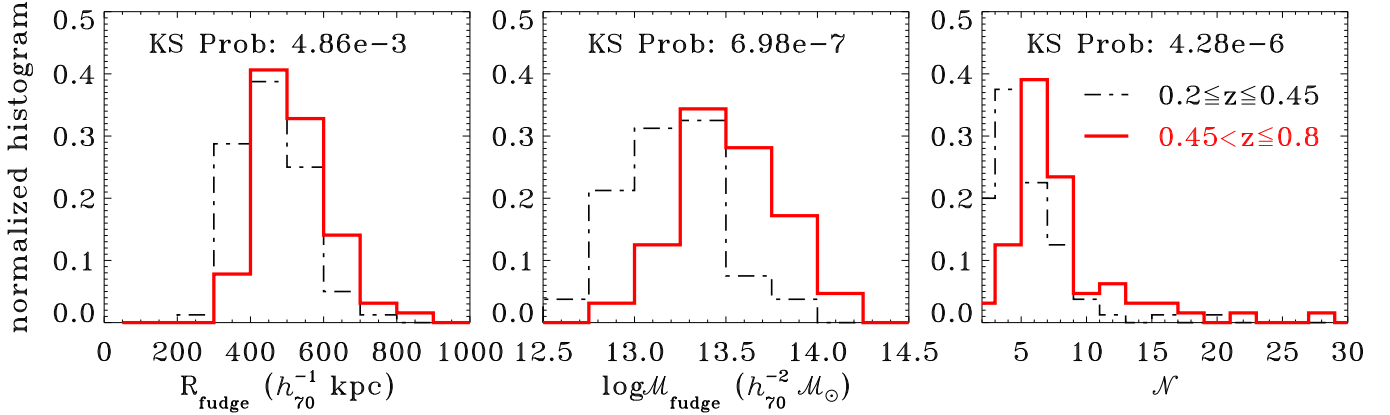}}
 \caption{Comparison of the general properties of groups in the
$0.2\leq z \leq 0.45$ (black dot-dashed line) and $0.45< z \leq 0.8$
(red solid line) redshift bin. From left to right we compare
$R_{fudge}$ (the estimate of the virial radius), $\log{\cal
M}_{fudge}$ (the estimate of the mass of the group), both fudge
quantities are calibrated with the mocks as defined in
\citet{Knobel2009}), and $\cal N$, as defined in Sect.
\ref{sect:vol_mass_lim_sample}.}
\label{fig:group_prop}
\end{figure*}

For each of the two redshift bins defined in the previous section,
$0.2\leq z \leq 0.45$ and $0.45< z \leq 0.8$, we proceeded to build
the corresponding \SG.  Notice that while for centering purposes we
used all spec-z and phot-z galaxies available in our group catalog,
irrespective of their mass and B-band rest-frame luminosity, for our
analysis we will use only galaxies within the mass volume-limited
samples as defined in Table \ref{tab:N_gals}.

Before moving to a detailed study of the group member galaxies
properties, it is interesting to compare the general properties of
groups in low- and high-redshift bins, to highlight any redshift-dependent 
trend in the group sample we used in our science
analysis.

In Fig.\ref{fig:group_prop} we compare from left to right
$R_{fudge}$ - the virial radius estimate \citep{Knobel2011},
$M_{fudge}$ - the mass of the group calibrated with the mocks as in
\citet{Knobel2009}, and the group richness, $\cal N$, as defined in
section \ref{sect:vol_mass_lim_sample}, for low- (black dot-dashed
line) and high- (red solid line) redshift galaxy groups.  The KS test
always rejects with more than 99.99\% confidence the hypothesis that
properties of low and high redshift groups are drawn from the same
distribution. In the low-redshift bin on the mean we deal with
smaller, less massive and poorer groups than those in the highest-
redshift bin. This is not an unexpected result given that zCOSMOS is a
flux-limited survey and therefore the observed population of both
galaxies and groups varies with increasing redshift. As a consequence,
the group detection works only on progressively brighter/more massive
galaxies moving to higher redshifts. We shall need to take into
account these differences when discussing our results. We define a
subset of richer groups for the low-redshift bin using richness $\cal
N$, defined for this bin as the number of member galaxies surviving
the evolving magnitude cut-off: $M_{\rm cut-off} = M^*_{B {\rm ev}} +
2.1$ (see Sect. left panel of Fig.\ref{fig:Mag_z_plus_UB_mass_dist}).
We adopted a separation of ${\cal N} \leq (>) 12$ to distinguish
between poor(rich) groups, a value roughly corresponding to
$M_{fudge} \leq (>) 13.3$, so that rich low-z groups are virtually
indistinguishable in mass distribution from the high-redshift
sample. Indeed, while a KS test comparing the distributions of
$M_{fudge}$ of poor and rich groups defined this way rejects the 
hypothesis that they are drawn from the same distribution with more
than 99.99\% confidence , the KS test comparing distributions of $M_{fudge}$
of rich low-z groups and of high-z groups does not reveal
 any significant difference between the two.

\begin{figure*}[h!bt]
\centering
\resizebox{\hsize}{!}{\includegraphics[bb=61 390 528 620]{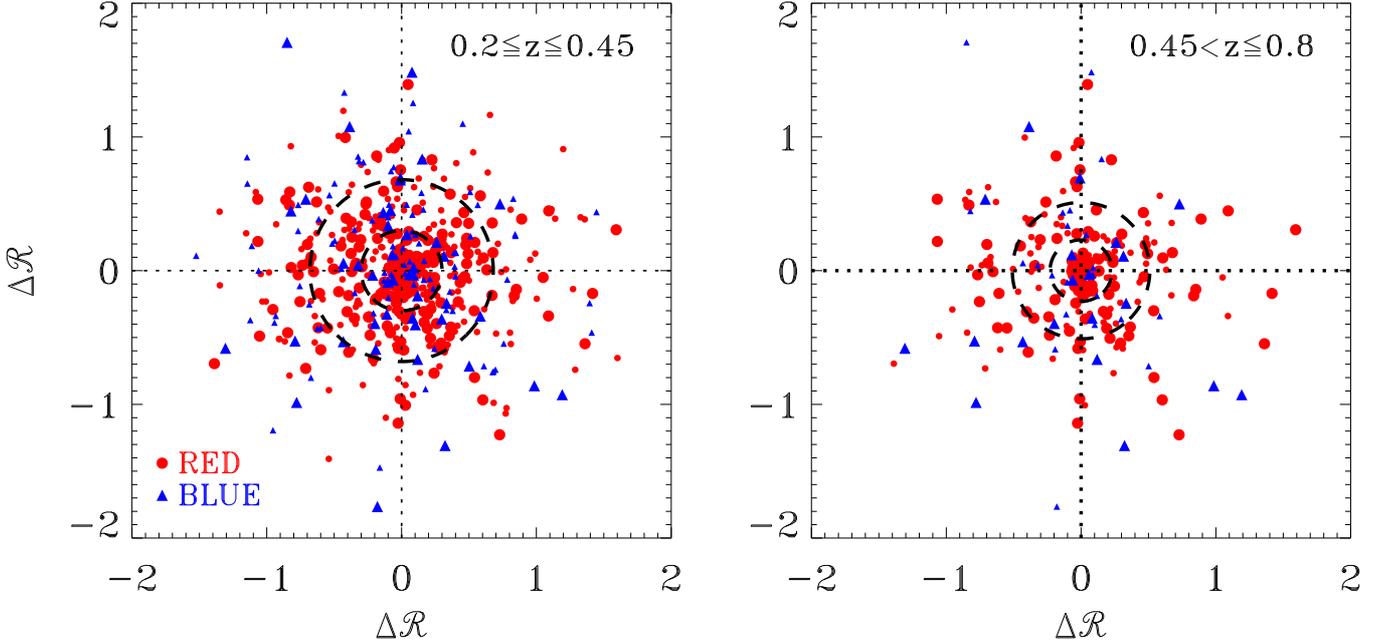}}
\caption{Sky distribution of the galaxies belonging to the low-z
(left) and high-z (right) composite group. Ra-dec positions are
expressed in terms of the rescaled distances ${\cal R}$. Points are
colored according to the $(U-B)$ colors of the galaxies, while point
dimensions are scaled according to the masses of the galaxies.  As a
reference we draw dashed circles corresponding to the different
central/intermediate/external region limits in each composite group.}
\label{fig:sky_plot_composite}
\end{figure*}

To explore how galaxy population properties change as a function of
group-centric distance, we first sorted \SG\, galaxies into increasing
scaled distances from \SG\, center and then divided their distribution
into three equipopulated bins corresponding to inner, intermediate, and
peripheral \SG\, regions.

In Table \ref{tab:shell_range} we list the exact radial ranges of each
of these three regions, all values are normalized to $R_{fudge}$.  The
corresponding three median distances are ${\cal R} \sim 0.15$, ${\cal
R} \sim 0.4$ and ${\cal R} \sim 0.85$, respectively, therefore these
regions can be considered as the group inner core, intermediate, and more
external/in-fall region.

Given that the median $R_{fudge}$ is $\sim 500$ \Kpch\, in both
redshift bins, the inner region extends typically up to $\sim 150$
\Kpch. Because the VW center is on the average only $\sim 40$ \Kpch\,
away from the group fiducial center (see Sec. \ref{sect:center}), our
error in centering is negligible with respect to the median inner
region size, and should not have a significant impact when exploring
the group-centric dependence of galaxy properties.

\begin{table*}
\caption{Radial range explored in the three \SG\, regions. All
distances ${\cal R}$ are normalized to $R_{fudge}$.}
\label{tab:shell_range}
\centering
\begin{tabular}{l c c c c c c}
\hline
 Redshift &  \multicolumn{2}{c}{$1^{st}$ region}  & \multicolumn{2}{c}{$2^{nd}$ region}  & \multicolumn{2}{c}{$3^{rd}$ region}   \\
  & range & ${\cal R}_{median}$ & range & ${\cal R}_{median}$ & range & ${\cal R}_{median}$ \\
\hline
 $0.2 \leq z \leq 0.45$ & ${\cal R} \leq 0.30$ & 0.15 & $0.30 < {\cal R}\leq 0.68$ & 0.47 & ${\cal R} > 0.68$ & 0.94 \\
 $0.45 < z \leq 0.8$ & ${\cal R} \leq 0.23$ & 0.13 & $0.23 < {\cal R}\leq 0.51$ & 0.37 & ${\cal R} > 0.51$ &  0.74 \\   
\hline
\end{tabular}
\end{table*}

The sky distribution of galaxies belonging to the low-z (left) and
high-z (right) composite group is shown in
Fig.\ref{fig:sky_plot_composite}.  Points are coded according to the
$(U-B)$ colors of the galaxies, while point dimensions are scaled
according to galaxy masses. As a reference we draw dashed circles
corresponding to the division between the different regions in each
composite group. We note that the overall shape of the composite group
has a well-defined peak corresponding to the center, while the
projected density decreases as we move from the center to the
outskirts. A visual inspection of the galaxy sky distribution
already shows rough differences in masses and colors depending on the
area we explore.  In the next section we proceed to extensively
analyze these trends and their dependence on intrinsic galaxy/group
properties, properly accounting for possible field contamination
effects.

\section{Results} 
\label{sect:Results}

We will start our analysis by exploring how galaxy colors are affected
by group environment, irrespective of galaxy position within the group
(see Sect.\ref{sect:global_trends}).  We will then move to investigate
the presence of color segregation within the group environment
(Sect.\ref{sect:color_seg}), and if the effect of the group
environment extends to scales somewhat larger than those of the group
size itself (see Sect.\ref{sect:near_far_field}).  Thanks to the high
statistic of the \20K we will also be able to investigate if and how
observed trends depend on group richness and on galaxy stellar
mass (see Sect.\ref{sect:color_seg_details}). Finally we will search
for evidence of mass segregation inside groups and how it might depend
on group richness and affect observed mass trends (see
Sect.\ref{sect:mass_seg}).

\begin{figure*}
 \centering
  \resizebox{\hsize}{!}{\includegraphics[bb=54 371 553 720]{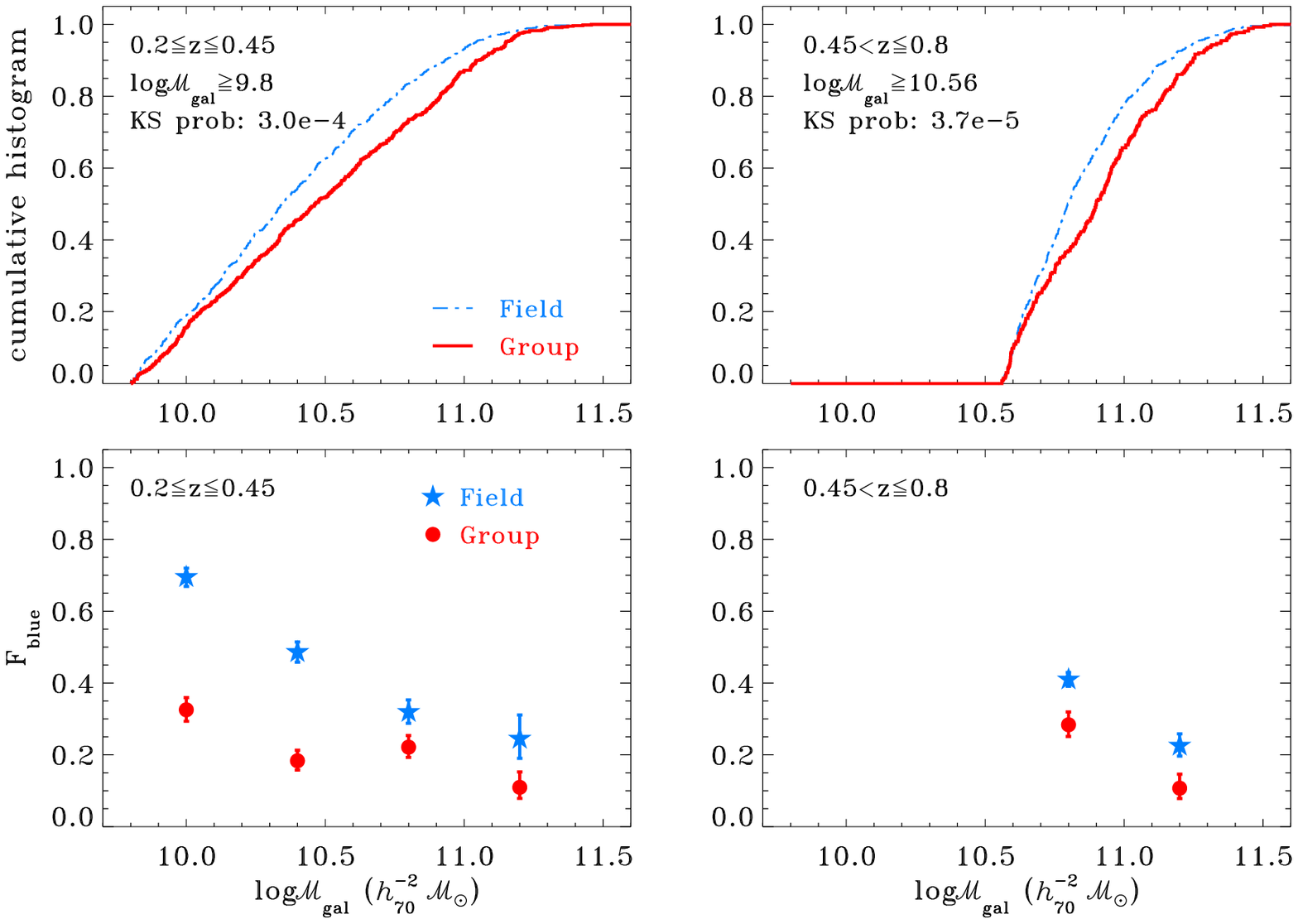}}
   \caption{{\it Top panels}: cumulative galaxy stellar mass
distribution of the mass-complete group and field samples in red and
cyan colors. {\it Bottom panels}: $F_{\rm blue}$ at fixed
galaxy mass in groups (red circles) and field (cyan stars).  {\it
Left panels} always refer to the low-redshift bin, while {\it right
panels} refer to the high redshift one. We used galaxy mass bins 0.4
dex wide.}
 \label{fig:mass_environment_joint}  
\end{figure*}

\subsection{$F_{\rm blue}$ and galaxy stellar masses in groups vs field}
\label{sect:global_trends}

The cumulative galaxy stellar mass distributions of the mass-complete
group and field samples are shown in the top panels of
Fig.\ref{fig:mass_environment_joint}, red solid and cyan dot-dashed
lines, respectively, the left(right) panels refer to the low(high)
redshift bin.  The KS test rejects the hypothesis that group and field
galaxy mass distributions are drawn from the same population with more
than 99.99\% confidence for both redshift bins. Group environment
hosts preferentially more massive galaxies than the field one,
confirming well-known literature results \citep{Iovino2010,
Kovac2010, Bolzonella2010}.

As a consequence, to explore the presence of color trends as a
function of environment, we need to separate the joint effect of mass
and environment and to perform the analysis in narrow mass bins of
galaxy stellar mass.  We adopted a galaxy stellar mass bin of 0.4 dex,
which is approximately twice our error in estimating galaxy stellar
masses \citep{Pozzetti2010}. Bottom panels of
Fig.\ref{fig:mass_environment_joint} show $F_{\rm blue}$ at fixed
galaxy stellar mass in groups (red circles) and field (cyan stars),
while Table \ref{tab:F_blue_mass_bins} lists in detail the $F_{\rm
blue}$ values and their errors. Both at high and low redshift, the
blue fraction increases when moving toward less massive
galaxies. $F_{\rm blue}$ is always higher in the field than in the group, a
difference that decreases moving to more massive galaxies. The most
massive galaxies ($\log({\cal M}_{gal}/{\cal M_{\odot}}) > 11.0$) do not show any
significant $F_{\rm blue}$ evolution with redshift within the error
bars. For the galaxies with $10.6 \leq \log({\cal M}_{gal}/{\cal M_{\odot}}) \leq 11.0$,
$F_{\rm blue}$ decreases for both group and field environment when
moving from high to low redshift. 

Notice that on average, groups in the low-z
bin are poorer than those in the high-z bin. As we will show below, 
see Sect.\ref{sect:color_seg_details}, $F_{\rm blue}$ depends on the 
group richness, $F_{\rm blue}$ being lower in richer groups.
As a consequence, the decrease of $F_{\rm blue}$  across the explored 
redshift range for the sample of group galaxies should be even more 
pronounced.

\begin{table}[h] 
\caption{Observed blue fractions in groups and field for different
galaxy stellar mass bins.}
\label{tab:F_blue_mass_bins}
\centering
\begin{tabular}{l c c}
\hline
 Sample $0.2\leq z\leq 0.45$ &  group  & field  \\
\hline
  &   &        \\
 $9.8 \leq \log({\cal M}_{gal}/{\cal M_{\odot}})  \leq 10.2$   &  $0.33^{+0.03}_{-0.03}$ & $0.70^{+0.03}_{-0.03}$     \\
   &   &       \\                                           		        	   			   
 $10.2 \leq \log({\cal M}_{gal}/{\cal M_{\odot}})  \leq 10.6$  &  $0.18^{+0.03}_{-0.03}$ & $0.49^{+0.03}_{-0.03}$     \\
   &   &      \\                                            		        	   			      
 $10.6 \leq \log({\cal M}_{gal}/{\cal M_{\odot}}) \leq 11.0$   &  $0.22^{+0.03}_{-0.03}$ & $0.32^{+0.03}_{-0.03}$   \\
   &   &     \\                                             		        	   			     
 $11.0 \leq \log({\cal M}_{gal}/{\cal M_{\odot}}) \leq 11.4$   &  $0.11^{+0.04}_{-0.03}$ & $0.25^{+0.07}_{-0.06}$   \\
   &   &     \\
\hline
 Sample $0.45< z\leq 0.8$  &  group   & field  \\
\hline
   &   &      \\        
 $10.6 \leq \log({\cal M}_{gal}/{\cal M_{\odot}}) \leq 11.0$  &  $0.28^{+0.04}_{-0.03}$ & $0.41 ^{+0.02}_{-0.02}$       \\
   &   &    \\                                              					   
 $11.0 \leq \log({\cal M}_{gal}/{\cal M_{\odot}}) \leq 11.4$  &  $0.11^{+0.04}_{-0.03}$ & $0.23 ^{+0.03}_{-0.03}$       \\
   &   &     \\
\hline
\end{tabular}
\end{table}

At fixed galaxy stellar mass, the migration to the red sequence happens
earlier in the groups and later in the field, suggesting the presence
of physical mechanisms able to remove gas that causes the earlier
quenching of galaxies in groups. We remind the reader that any
contamination of the group sample by field galaxies and vice versa, for
which we are not applying any correction, will only render the observed 
trends less prominent. The real, corrected, trends therefore
would be even more pronounced. This result excellently agrees
with our previous zCOSMOS results on groups \citep{Iovino2010,
Bolzonella2010, Peng2010}.

The questions we will address in the following sections are: do the
group member galaxies all share the same $F_{\rm blue}$ value
irrespective of their position within the group? Do the galaxies
located in the central region of groups share the same mass
distribution as the galaxies located in the group outskirts?
Ideally, the first question is better addressed in narrow bins of
galaxy stellar mass to avoid the mass-color degeneracy. However, even
with such a dataset as the \20K, we are still limited by small number
statistics, when splitting our sample according to distances from group
center and therefore we will start our analysis
working in cumulative mass ranges in the next section.

\begin{figure*}
 \centering
  \resizebox{\hsize}{!}{\includegraphics[bb=54 388 542 606]{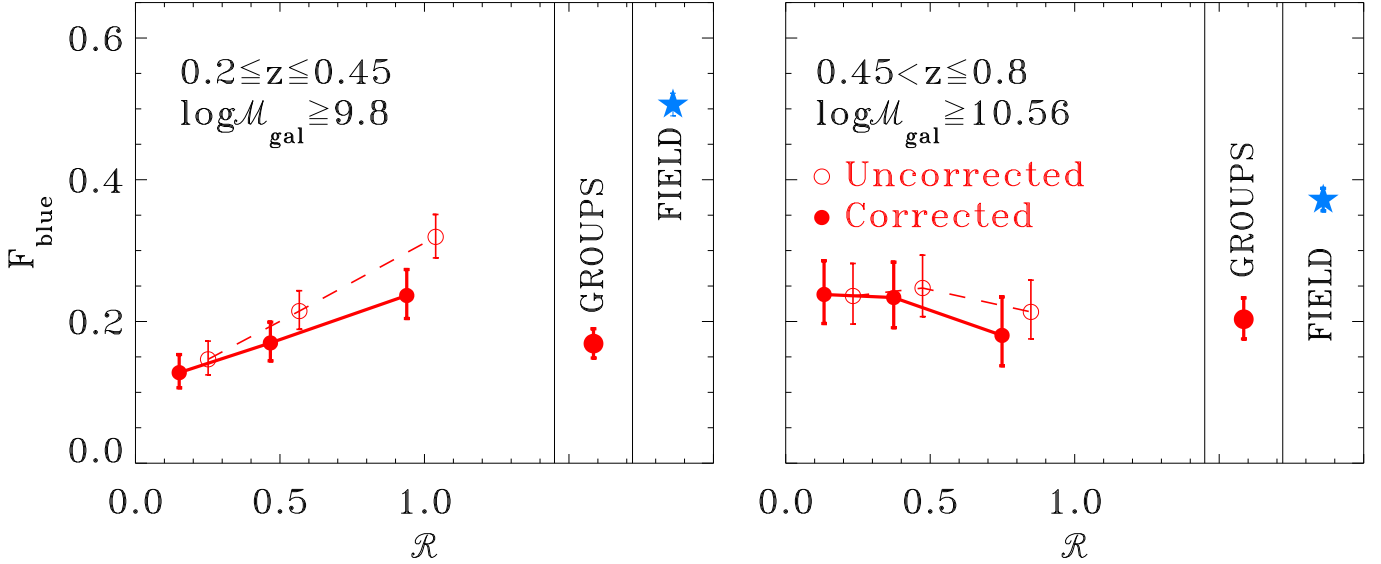}}
 \caption{\SG\, blue fraction as a function of the group-centric
 distance.  Left/(right) panels refer to the low/(high) redshift bin
 and mass limits are ${\cal M}_{\rm cut-off}=9.8$/(${\cal M}_{\rm
 cut-off}=10.56$).  Open circles refer to observed ${F_{\rm blue}}$
 while filled ones to corrected ${F_{\rm blue}}$, \ie taking into
 account interloper contamination (see text for details).  The points
 refer to the three regions: inner core, intermediate, and more
 external/infall region. Corrected blue fractions values are displayed
 at the median distance of each region, while observed blue fractions
 are offset for clarity. As a reference we plot the fraction of
 the blue field galaxies (cyan star) and the mean corrected blue
 fraction in groups (big red circle).  
 }
 \label{fig:F_blue}
\end{figure*}

\subsection{Color segregation: $F_{\rm blue}$ as a function of the group-centric distance}
\label{sect:color_seg}

To explore the color segregation, we measured $F^{GR}_{blue,obs}$, the
observed fraction of blue group member galaxies, in each of the three
\SG\, regions defined in section \ref{sect:two-super}. We used the
formula

\begin{equation}
F^{GR}_{blue,obs}=\frac{N^{GR}_{blue,obs}}{N^{GR}_{tot,obs}},
\end{equation}

where the index GR refers to group galaxies.
However, to establish the reliability of any observed changing
mix of galaxy colors at different radial distances from the group
center, it is essential to properly account for the presence of field
galaxy contaminants and their (changing) relative contribution at
differente distances from the group center, because the field population
has a higher $F_{\rm blue}$ than the group.  We thus need to
estimate the corrected group fraction of blue galaxies, $F^{GR}_{\rm
blue}$:

\begin{equation}
F^{GR}_{\rm blue}=\frac{N^{GR}_{\rm blue}}{N^{GR}_{\rm tot}}=\frac{N^{GR}_{blue,obs}-N^{INT}_{blue}}{N^{GR}_{tot,obs}-N^{INT}_{tot}},
\label{eq:f_blue}
\end{equation}

that is, the corrected ratio of the blue member galaxies to the total
number of member galaxies, after excluding the percentage of
interlopers - indicated by the index $INT$ - entering in the observed
list of group members because of group detection algorithm failures.

If $\mathcal{PI}$ is the estimated percentage of interlopers defined
in Sect.\ref{sect:how_build_sup}, we can obtain $N^{INT}_{tot}$ for
each group region by $N^{INT}_{tot} = {\mathcal PI} \times
N^{GR}_{tot,obs} = {\mathcal PI} \times (N^{GR}_{tot} +
N^{INT}_{tot})$. The value of $N^{INT}_{blue}$ can simply be estimated
as $N^{INT}_{blue} = F^{FIELD}_{blue} \times N^{INT}_{tot}$, where
$F^{FIELD}_{blue}$ is the known blue fraction of the field
galaxies. This way we take into account the radial trend of interloper
contamination and have all ingredients to retrieve the
corrected values of $F^{GR}_{\rm blue}$ from equation \ref{eq:f_blue}.

In Fig.\ref{fig:F_blue} the left/(right) panel shows the blue fraction
as a function of the group-centric distance in the \SG\, for the
low/(high) redshift bin in the mass-complete sample down to ${\cal
M}_{\rm cut-off}=9.8$/(${\cal M}_{\rm cut-off}=10.56$).  Uncorrected
and corrected blue fraction values are indicated with open and filled
circles, respectively. Corrected blue fractions are displayed at the
median normalized ${\cal R}$ distance of galaxies in each region,
while uncorrected blue fractions are slightly offset for clarity.
As a reference we plot the fraction of blue field galaxies (cyan
stars) and of the whole group (big red filled circle, corrected values
only). Error bars are estimated using the approximate analytical
formulas for a binomial distribution provided by
\citet{Gehrels1986}. In Table \ref{tab:F_blue} we list the values of
observed and corrected ${F_{\rm blue}}$ for each region and sample
considered.

\begin{table*}
\caption{Blue fractions in groups, observed and corrected for field
contamination. Last two columns list total group and field values.}
\label{tab:F_blue}
\centering
\begin{tabular}{l c c c c c c c c}
\hline
 Sample $0.2\leq z\leq 0.45$ &   \multicolumn{2}{c}{$1^{st}$ region}  &  \multicolumn{2}{c}{$2^{nd}$ region}  &  \multicolumn{2}{c}{$3^{rd}$ region}  & group    & field  \\
                             &  obs & corr & obs & corr & obs & corr & corr  &   \\
\hline
  &   &   &    &    &   &    &   &  \\
 $\log({\cal M}_{gal}/{\cal M_{\odot}}) \geq 9.8$  & $0.15^{+0.03}_{-0.02}$ & $0.13^{+0.03}_{-0.02}$ & $0.22^{+0.03}_{-0.03}$ & $0.17^{+0.03}_{-0.03}$ &  $0.32^{+0.03}_{-0.03}$ & $0.24^{+0.04}_{-0.03}$ & $0.17^{+0.02}_{-0.02}$ & $0.51^{+0.02}_{-0.02}$  \\
  &   &   &    &    &   &    &    & \\
 $\log({\cal M}_{gal}/{\cal M_{\odot}}) \geq 10.56$  & $0.11^{+0.04}_{-0.03}$ & $0.10^{+0.04}_{-0.03}$ & $0.16^{+0.04}_{-0.03}$ & $0.13^{+0.05}_{-0.03}$ &  $0.25^{+0.05}_{-0.04}$ &  $0.22^{+0.06}_{-0.05}$ & $0.14^{+0.03}_{-0.03}$ & $0.30^{+0.03}_{-0.03}$ \\   
  &   &   &    &    &   &    &    & \\
 $\log({\cal M}_{gal}/{\cal M_{\odot}}) \geq 9.8$ \& $5 \leq {\cal N} \leq 12$ & $0.18^{+0.04}_{-0.04}$ & $0.16^{+0.04}_{-0.04}$ & $0.20^{+0.04}_{-0.04}$ &  $0.15^{+0.04}_{-0.04}$ & $0.43^{+0.05}_{-0.05}$ & $0.39^{+0.06}_{-0.06}$ & $0.22^{+0.03}_{-0.03}$ & $0.51^{+0.02}_{-0.02}$ \\   
  &   &   &    &    &   &    &    & \\
 $\log({\cal M}_{gal}/{\cal M_{\odot}}) \geq 9.8$ \& ${\cal N} > 12$ & $0.14^{+0.04}_{-0.03}$ & $0.13^{+0.04}_{-0.03}$ & $0.23^{+0.04}_{-0.04}$ &  $0.19^{+0.04}_{-0.04}$ & $0.28^{+0.05}_{-0.04}$ & $0.18^{+0.05}_{-0.04}$ & $0.16^{+0.03}_{-0.03}$ & $0.51^{+0.02}_{-0.02}$ \\   
  &   &   &    &    &   &    &    & \\
\hline
 Sample $0.45< z\leq 0.8$  &   \multicolumn{2}{c}{$1^{st}$ region}  &  \multicolumn{2}{c}{$2^{nd}$ region}  &  \multicolumn{2}{c}{$3^{rd}$ region}  & group    & field  \\ 
                             &  obs & corr & obs & corr & obs & corr &   &  \\
\hline
  &   &   &    &    &   &    &   &   \\
 $\log({\cal M}_{gal}/{\cal M_{\odot}}) \geq 10.56$ &  $0.24^{+0.05}_{-0.04}$ & $0.23^{+0.05}_{-0.04}$ & $0.25^{+0.05}_{-0.04}$ & $0.23^{+0.05}_{-0.04}$ &  $0.21^{+0.05}_{-0.04}$ &  $0.18^{+0.05}_{-0.04}$ & $0.20^{+0.03}_{-0.03}$ & $0.37^{+0.02}_{-0.02}$ \\   
  &   &   &    &    &   &    &   &  \\
\hline
\end{tabular}
\end{table*}

In the low-redshift bin, the observed ${F^{GR}_{\rm blue}}$ rises as
the distance from the group center increases (left panel of
Fig.\ref{fig:F_blue}). Though becoming less prominent, this color
segregation holds even when correcting for interlopers
contamination. Considering the field point, we can see a trend of
increasing ${F_{\rm blue}}$ moving from the inner core of groups to
their outskirts and farther away to the field.

The difference between ${F^{GR}_{\rm blue}}$ in the inner core and
that in the outskirts is 1.7$\sigma_{1^{st}-3^{nd}}$, where
$\sigma_{1^{st}-3^{nd}}$ is the sum in quadrature of $\sigma_{1^{st}}$
and $\sigma_{3^{rd}}$, the error of the fraction of blue galaxies in
the inner core and outskirts. The difference between
${F^{GR}_{\rm blue}}$ in the outskirts and $F^{FIELD}_{blue}$ is
7$\sigma_{3^{nd}-field}$.  Entering the group potential well has a
considerable influence in changing galaxy colors, but the most relevant
difference is between ${F^{GR}_{\rm blue}}$ in the inner core and
$F^{FIELD}_{blue}$: 11$\sigma_{1^{st}-field}$.

Moving to the highest redshift bin, we do not detect any significant
color radial trend within the \SG, while we still detect a significant
difference with respect to the field galaxy population (right panel of
Fig.\ref{fig:F_blue}).  The group blue fraction is at a constant value
of ${F^{GR}_{\rm blue}} \sim 0.1$, a value 3.9$\sigma$ lower than the
corresponding blue fraction in the field sample. The values of
${F_{\rm blue}}$ for group and field are significantly lower in this
redshift bin than those obtained in the low redshift bin discussed
previously.  In order to understand this result, we need to remember
that we applied a higher galaxy stellar mass cut-off in the
high-redshift bin: $\log({\cal M}_{\rm gal}) \geq 10.56$, thus we are
observing galaxies more massive than in the low redshift bin, which
is the obvious cause for the general lowering of the values of
${F_{\rm blue}}$ for the field and the group population.  Another
factor to consider is that we are observing groups that are on average
more massive than their low-redshift counterparts (see
Fig. \ref{fig:group_prop}).

Before moving to investigate in detail how our findings depend on
group richness and galaxy stellar masses in
Sec. \ref{sect:color_seg_details}, we will explore in the next section
whether the observed differences between group and field stop at group
boundaries, or if there is a continuous trend of increasing $F_{\rm
blue}$ in the closest proximity of groups.

\begin{figure}
\centering
\resizebox{\hsize}{!}{\includegraphics[bb=54 388 327 614]{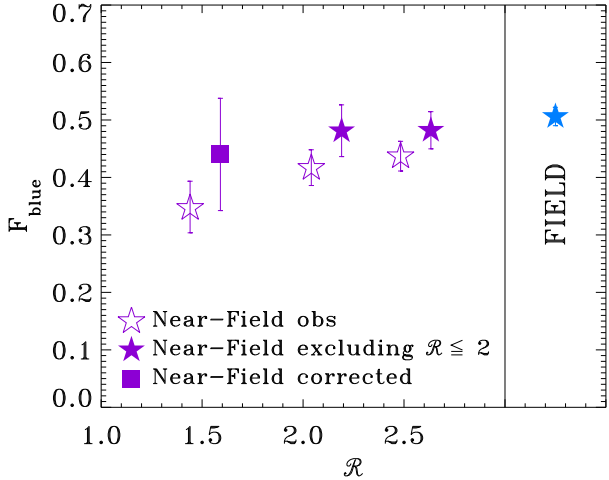}}
\caption{$F_{blue}$ of {\it near-field} galaxies (violet empty stars) as a
function of increasing projected radial distances from groups in the
low-redshift bin: ${\cal R} \leq 2.0$, ${\cal R} \leq 3.0$ and ${\cal
R} \leq 4.0$.  Violet filled stars show $F_{blue}$ of {\it near field}
after excluding {\it near-field} galaxies with projected radial
distances ${\cal R} \leq 2.0$.  The filled square shows $F_{blue}$ for
the first {\it near-field} annulus, after correcting for the 15\%
spectroscopic galaxy-group incompleteness, see text for details.
Observed $F_{blue}$ values are displayed at the median distance of
each {\it near-field} region, while corrected ones are slightly offset
for clarity. As a reference we display the $F_{blue}$ of the
field with a cyan star.}
 \label{fig:F_blue_field}
\end{figure}

\subsection{{\it Near-field} vs global field: is there a large-scale $F^{FIELD}_{blue}$ trend?}
\label{sect:near_far_field}

The physical scale, in terms of density or projected distances, over
which environment begins to set up the well known correlations with
galaxy star-formation rates, colors, and morphologies is a question that is
still open \citep{Kauffmann2004, Balogh2004, Blanton2006}.  We
explored the possible presence of large-scale trends of
$F^{FIELD}_{blue}$, with the aim of detecting \eg colors
redder than those of field sample for the galaxy
population located in the closest proximity of our groups.  As discussed in
Sect. \ref{sect:field_sample}, we defined a sample of field galaxies
ideally not affected by group environment, and a sample of so-called
{\it near-field} galaxies, \ie galaxies located in the closest
proximity of groups.

In the low-redshift bin, where the sample size enables us to perform this
analysis, we split the {\it near-field} into subsets of three nested
rings of increasing projected radial distances from the \SG\, center that were
defined as follows: ${\cal R} \leq 2.0$, ${\cal R} \leq 3.0$ and
${\cal R} \leq 4.0$, and measured $F_{\rm blue}$ for each of
them. These values are plotted in Fig.\ref{fig:F_blue_field} with
empty stars, and display a regular increase moving away from the group
center, progressively nearing the value obtained for the field sample.
However, this apparent continuous trend, extending beyond group size,
disappears when excluding field galaxies with projected radial
distances ${\cal R} \leq 2.0$ (filled stars) from each of the points shown in
Fig.\ref{fig:F_blue_field}, suggesting that the
trend in question is caused by contamination from missed group
members, predominantly located in the nearest neighborhood of groups.

The same result can be obtained correcting the values of the first
{\it near-field} ring with a procedure similar to that used for group
galaxies. The galaxy success rate of the group-finding algorithm turns
into a spectroscopic incompleteness on galaxy-group basis of 15\%, as
estimated in \citet{Knobel2009}. Correcting for this percentage of
contamination by group galaxies is enough to raise the observed value
of $F^{FIELD}_{blue}$ in the first {\it near-field} annulus to that of
the field, as shown by the filled square in
Fig.\ref{fig:F_blue_field}.

This analysis confirms that the physical scale on which the
environment plays its role coincides with the group physical scales,
in agreement with \citet{Kauffmann2004}, \citet{Blanton2006}, 
and \citet{Wilman2010}. It
therefore suggests that there is no transition region from field to
group domain and field galaxies start to be affected by the group
environment when they enter it.

Interstingly, it also implies that the values we estimated for the
incompleteness of our group catalog is realistic, because it produces, once
the corresponding correction is applied, values of $F_{\rm blue}$ for
the first {\it near-field} annulus that agree well with
field values.

\begin{figure*}
\centering
\resizebox{\hsize}{!}{\includegraphics[bb=54 388 542 606]{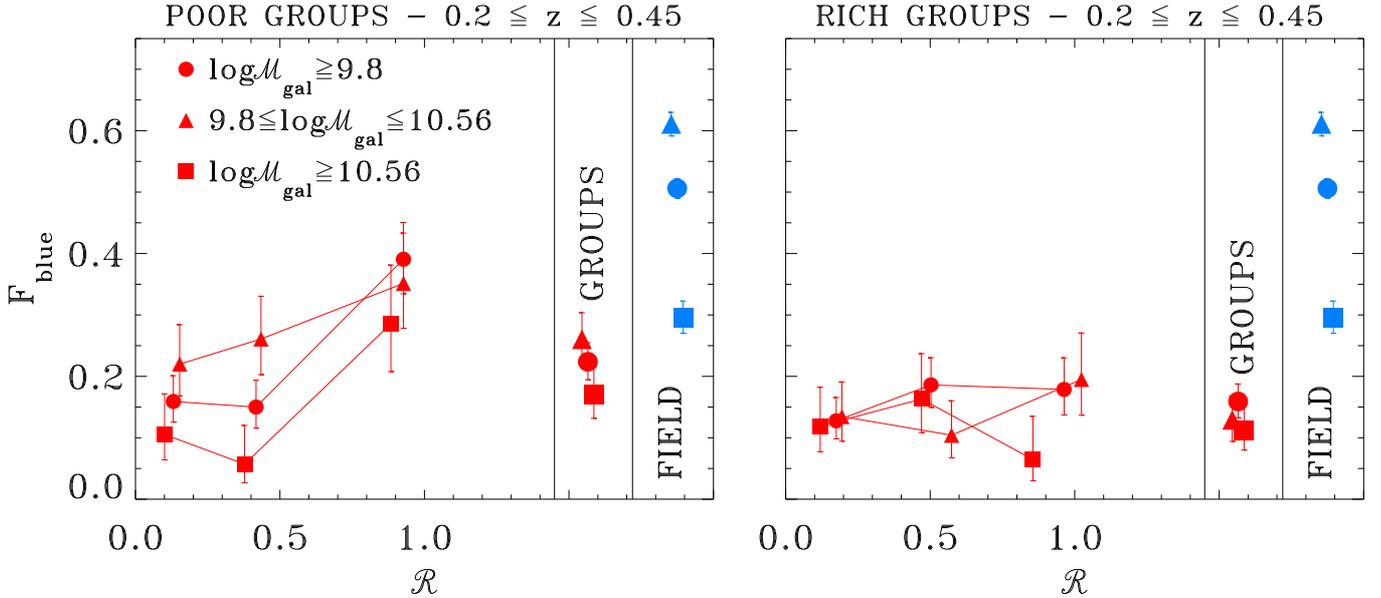}}
\caption{Corrected blue fraction as a function of the group-centric
distance in the low-z poor groups (left) and rich groups (right).  We
define poor(rich) groups as those with number of member galaxies
$\leq(>)12$ after applying the evolving magnitude cut off $M_{\rm
cut-off} = M^*_{B {\rm ev}} + 2.1$ (see
Fig.\ref{fig:Mag_z_plus_UB_mass_dist}). The mass bins adopted are
indicated in the legend.  Corrected blue fractions are displayed at
the median of the distances from the center for galaxies in each
region.  As a reference we also plot the fraction of the blue field
galaxies and the mean corrected blue fraction in groups. }
 \label{fig:F_blue_mass_rich}
\end{figure*}

\subsection{Color segregation: a closer look at galaxy mass and group richness dependencies}
\label{sect:color_seg_details}

In Sect.\ref{sect:color_seg} we have shown that low- and high-z group
galaxies display different radial color trends.  However, we also
noticed two important differences: the galaxy stellar mass cut-off
adopted ($\log({\cal M}_{gal}/{\cal M_{\odot}})\geq 9.8$ to be compared to 
$\log({\cal M}_{gal}/{\cal M_{\odot}}) \geq 10.56$), and the different range in group richness
spanned in these two redshift bins.  It is therefore interesting to
investigate how the observed radial trends of ${F_{\rm blue}}$ depend
on these two quantities. We split the low-z group sample according
to richness $\cal N$ defined as the number of member galaxies
surviving after applying the evolving magnitude cut-off $M_{\rm
cut-off} = M^*_{B {\rm ev}} + 2.1$, adopting a separation of ${\cal N}
\leq (>) 12$ to distinguish between poor(rich) groups.  We then
split the total galaxy sample for each of the two rich and poor
\SG\, into two bins of galaxy stellar mass: galaxies with stellar
masses $9.8 \leq \log({\cal M}_{\rm cut-off}) \leq 10.56$ and $\log({\cal
M}_{\rm cut-off}) \geq 10.56$.

In Fig. \ref{fig:F_blue_mass_rich} we show how $F_{\rm blue}$ varies
for the subsamples defined this way (from now on we display only
corrected values).  Poor(rich) groups are on the left(right) panel
 and triangles refer to lower stellar mass bin, squares to
higher stellar mass bin, and filled circles to the total mass
volume-limited sample. As a reference and with the same symbols for
each stellar mass bin, we plot the values of $F_{\rm blue}$ for field
galaxies and those for the total poor and rich group galaxies sample
considered.
 
The value of $F_{\rm blue}$ increases moving from higher to lower
galaxy stellar masses for each environment considered (see also
Sect.\ref{sect:global_trends}). However, there is a another trend
superimposed to this one: at fixed stellar mass the mean ${F_{\rm
blue}}$ is higher in poor groups than in rich ones, confirming
previous tentative results \citep[see][and references
therein]{Margoniner2001, Gerke2007, Iovino2010}.

Fig. \ref{fig:F_blue_mass_rich} shows for radial trends that bluer 
galaxies are preferentially located in the group outskirts only for 
poor groups . No such trend is observed for richer groups.  
For poor groups, galaxies with lower stellar mass show a
continuous trend of increasing $F_{\rm blue}$, whereas $F_{\rm blue}$
for most massive galaxies increases only in the outermost 
group region. Rich groups do not show any obvious radial trend.

Thus poor groups display higher $F_{\rm blue}$ values for their member
galaxies than richer groups and stronger radial trends at fixed galaxy
stellar mass than richer groups.  The observed galaxy color radial
trends become more evident moving from richer to poorer groups and
moving from higher to lower galaxy stellar masses.

We can therefore better explain the observed differences between low-
and the high-z groups trends discussed in Sect.\ref{sect:color_seg} as
caused by the different cut-off in galaxy stellar masses and to the 
different group richness ranges observed in the two different redshift
bins.

It is somewhat expected that at high-z the
massive groups we explore do not display any radial trend for the
massive galaxies shown in plot Fig.\ref{fig:F_blue}.

We will now proceed to explore if the differences observed in radial
trends between poor groups and rich groups relate to the possible
presence of galaxy stellar mass segregation within groups, a factor
that could be important in creating and/or enhancing observed color
trends.

\begin{figure*}
 \centering
  \resizebox{\hsize}{!}{\includegraphics[bb=54 388 542 606]{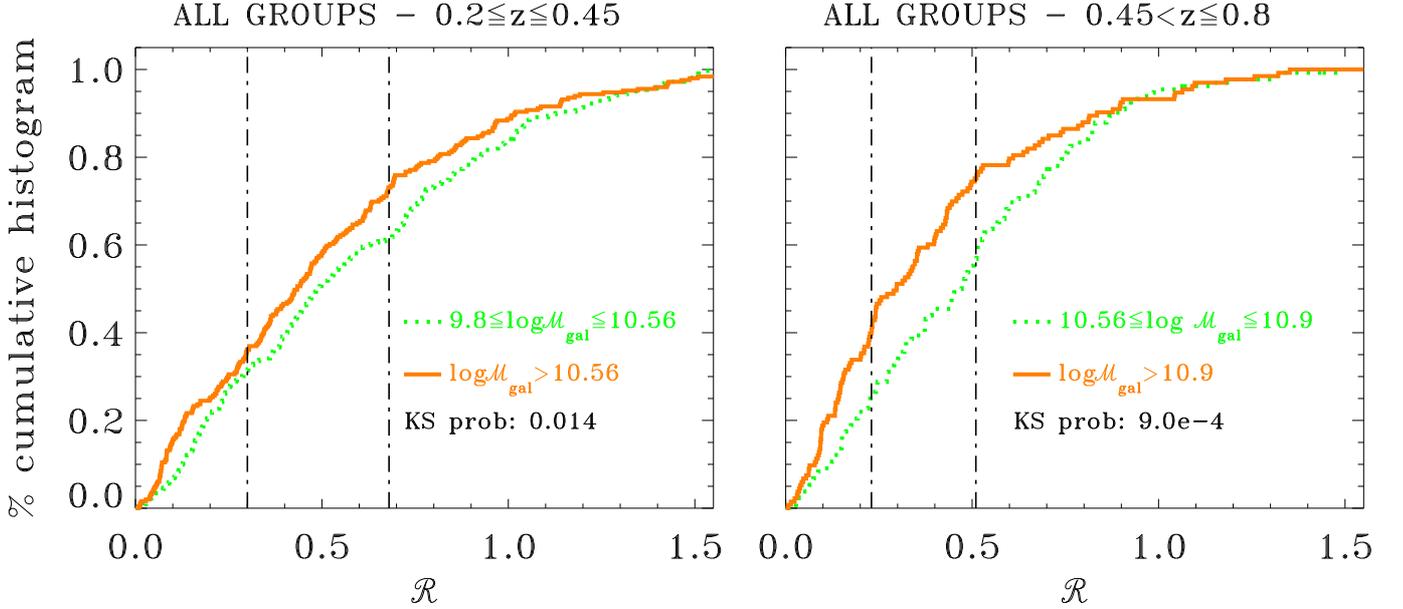}}
 \caption{Cumulative radial distribution of the galaxies belonging to
 each mass bin for both the lowest (left) and highest (right) redshift
 bin. We show the most massive galaxies with an orange solid line and the 
 less massive galaxies with a green dotted line. 
 As a reference we draw the limits of the
 first and second regions with a dot-dashed black line.}
 \label{fig:radial_dist_cumul_bothzbin}
\end{figure*}

\subsection{Mass segregation}
\label{sect:mass_seg}

The goal of this section is to check for the presence of mass segregation
within our group sample, to clarify if any of galaxy colors radial
trends we observed are simply the reflection of the galaxy colors and
stellar mass correlation coupled with varying galaxy stellar mass
functions moving from central to peripheral group regions.

We split the mass-complete galaxy sample into two stellar mass
bins: at low redshift the stellar mass limits chosen are $9.8 \le
\log({\cal M}_{gal}/{\cal M_{\odot}}) \le 10.56$ and $\log({\cal M}_{gal}/{\cal M_{\odot}}) > 10.56$, with a
total of 320(251) galaxies in the lowest(highest) mass bin. At high
redshift the stellar mass limits chosen are $10.56 \le \log({\cal
M}_{gal}/{\cal M_{\odot}}) \le 10.9$ and $\log({\cal M}_{gal}/{\cal M_{\odot}}) > 10.9$; in this case there
are 132(133) galaxies in the lowest(highest) mass bin.  In the
left(right) panel of Fig.\ref{fig:radial_dist_cumul_bothzbin} we show
the cumulative radial distribution of the galaxies belonging to these
bins for the low(high) redshift \SG s. The dotted line always refers
to the lowest stellar mass bin, while the solid line refers to the
highest stellar mass bin.  As a reference a dot-dashed black line
indicates the boundaries of the different group regions defined in
Sect.\ref{sect:two-super}.  In both redshift bins the most massive
galaxies preferentially populate the innermost regions, while the less
massive galaxies prefer the outer ones. A KS test confirms  the existence 
of a mass segregation for the low(high) redshift bin with
confidence higher than 98.6\%(99.99\%).

\begin{figure*}
 \centering
  \resizebox{\hsize}{!}{\includegraphics[bb=53 390 540 602]{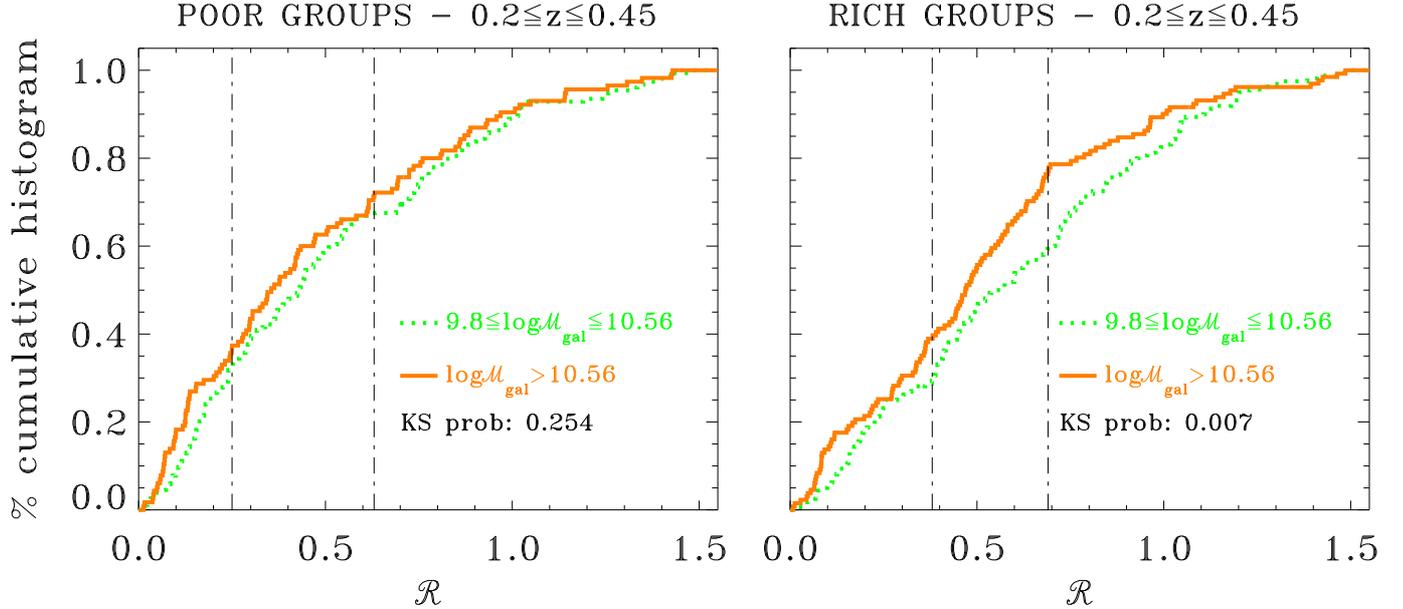}}
 \caption{Cumulative radial distribution of the galaxies belonging to
 each mass bin for poor (left) and rich (right) groups in the lowest
 redshift bin. We show the most massive galaxies with an orange solid line and the less
 massive galaxies with a green dotted line. As a reference we draw
 the limits of the first and second regions with a dot-dashed
 black line.}
 \label{fig:radial_dist_cumul_rich}
\end{figure*}

We here also explored the group richness dependency of the mass
segregation by repeating the same radial analysis but dividing the
low-z bin groups into poor and rich subsamples (see Section
\ref{sect:color_seg} for definitions).  The cumulative radial
distributions of galaxies are plotted in
Fig. \ref{fig:radial_dist_cumul_rich} for poor groups in the left panel, and for rich
groups in the right panel. In both panels dotted lines refer to galaxies
with $9.8 \le \log({\cal M}_{gal}/{\cal M_{\odot}}) \le 10.56$, while the solid 
lines denote galaxies at $\log({\cal M}_{gal}/{\cal M_{\odot}}) > 10.56$.  
For poor groups there is no
significant mass segregation because the KS test results are consistent
with the hypothesis that the radial distribution of galaxies from the
two mass bins are drawn from the same distribution. Vice versa,
galaxies located in rich groups display a significant mass
segregation, with most massive galaxies being closer to the \SG\,
center than the less massive ones. In this case the KS test rejects
the hypothesis that the radial distribution of galaxies from the two 
mass bins are drawn from the same distribution with more than 99.3\% confidence.

We performed a KS test comparing the galaxy stellar mass distribution in
different \SG\, regions and field galaxies.  For low-z rich and high-z
groups there is no significant difference between field galaxies and
the outermost \SG\ region, while the inner/intermediate regions
display a significant difference with respect to the field, in
agreement with their observed galaxy stellar mass segregation.  For
low-z, poor groups the galaxy stellar mass distribution is only
marginally different from that of field galaxies (a $\sim 2.5\sigma$
result), and this outcome holds irrespective of group-centric
distance, in agreement with the absence of galaxy stellar mass
segregation in these groups. In Table \ref{tab:shells_vs_field_mass} we
detail all numerical results of the various KS tests.

\begin{table*}
\caption{KS test probabilities that the mass distribution of the
galaxies in each group region and that of the field sample of galaxies
are drawn from the same distribution. Redshift ranges and group
richness are indicated in the Sample column.}
\label{tab:shells_vs_field_mass}
\centering
\begin{tabular}{l c c c}
\hline
 Sample &  $1^{st}$ gr region vs field &  $2^{nd}$ gr region vs field & $3^{rd}$ gr region vs field \\
\hline
 $0.2\leq z\leq 0.45$ \& $\log({\cal M}_{gal}/{\cal M_{\odot}}) \geq 9.8$ & $5.7\times10^{-4}$ & $1.1\times10^{-4}$ &  0.176  \\
 $0.2\leq z\leq 0.45$ \& $\log({\cal M}_{gal}/{\cal M_{\odot}}) \geq 9.8$ \& ${\cal N} \leq 12$  & 0.012 & 0.044  & 0.067 \\
 $0.2\leq z\leq 0.45$ \& $\log({\cal M}_{gal}/{\cal M_{\odot}}) \geq 9.8$ \& ${\cal N} > 12$ & $9.1\times10^{-4}$  & $1.3\times10^{-3}$ &  0.425   \\
 $0.45< z\leq 0.8$ \&  $\log({\cal M}_{gal}/{\cal M_{\odot}}) \geq 10.56$ & $6.8\times10^{-9}$  & 0.014  &   0.841 \\
\hline
\end{tabular}
\end{table*}

One could argue that the mass segregation we are detecting within rich
low-z groups and high-z groups increases through interloper
contamination. On average, interlopers would preferentially populate
group peripheral regions, where we actually observe a galaxy stellar
mass distribution that resembles more that of field galaxies, thus
producing the observed stellar galaxy mass distribution radial trends.
We used a Monte Carlo technique to establish the robustness
of our results with respect to this effect.  For
each of the two \SG s - rich low-z and high-z - we depleted the
mass-complete sample of its galaxy members by the estimated interloper
fraction in each group region. We used galaxy colors to select the
most probable interlopers, so that the galaxies removed had a value of
$F_{\rm blue}$ equal to that estimated for field galaxies (see
Sect.\ref{sect:global_trends}).  We performed this exercise 1000
times, keeping constant the total number of galaxies (\ie randomly
counting some of the surviving galaxies twice). Each time we estimated
the KS test probability that the radial distribution of less and most
massive member galaxies were drawn from the same radial distribution.

For rich low-z groups the KS test confirms a mass segregation with a
median confidence level of $97.8_{-5.9}^{+1.9}$.  For high-z groups,
the KS test confirms a mass segregation with a median confidence level
of $99.9_{-1.2}^{+0.1}$. In both cases the quoted errors correspond to
the lowest and highest quartiles of the KS test probability
distribution. We can conclude that at high-z the signal for a genuine
and significant radial stellar mass segregation for group galaxies is
strong and reliable, while at low-z the observed mass segregation for
rich group galaxies is somewhat enhanced by interloper contamination,
but, albeit at lower significance, an indication of its existence is
still present.

The main result of this section is therefore that low-z poor groups do
not show significant mass segregation; whereas low-z rich groups and
high-z groups, whose group mass ranges are somewhat similar, display a
significant mass segregation in the galaxy stellar mass ranges
explored. 

Because low-z poor groups are those that show a significant color
segregation, which is undetected in low-z rich groups and high-z
groups (see Sect.\ref{sect:color_seg_details}), we conclude that the
changing mix of color and masses are unrelated phenomena, possibly
originating from different physical mechanisms.  In the next section
we will discuss a possible interpretation.

\section{Migration from blue to red: the effects of group environment}
\label{sect:discussion}

Our analysis has confirmed that stellar mass is an important parameter
in the discussion of environmental influence on galaxy properties
and their evolution.

In Sect.\ref{sect:global_trends} we have shown that for massive
galaxies, $\log({\cal M}_{gal}/{\cal M_{\odot}}) \geq 11.0$, the value of $F_{\rm blue}$ does
not change moving from field to group galaxies: most massive galaxies
are red and {\it dead} irrespective of the environment they live in, a
well-known result both at low redshift \citep[see \eg][and references
therein]{Hansen2009, Bamford2009, Kimm2009} and at intermediate/high
redshift \citep[see \eg][and references therein]{Iovino2010,
Kovac2010, Peng2010, McGee2010}. 

Below this mass threshold we observe a gradual 'opening up' of the
difference between group and field values, so that lower values of
$F_{\rm blue}$ are reached earlier in groups than in the field (see
Fig.\ref{fig:mass_environment_joint}), confirming results previously
obtained using the zCOSMOS \10K sample \citep[see \eg][and references
therein]{Iovino2010, Kovac2010, Bolzonella2010} and recent 
results from the COSMOS survey \citep{George2011}. 

In addition to these trends, our analysis has shown that moving to masses below $\log({\cal
M}_{gal}/{\cal M_{\odot}}) = 11.0$ some subtler differences emerge when observing trends
within groups, both as a function of galaxy stellar mass and as a
function of group richness.  In the following discussion we
concentrate on the lower redshift bin of our sample ($0.2 < z <
0.45$), where we were able to study in detail the joined effect of
galaxy stellar mass and group richness on galaxy colors.  The
conclusions we will infer are easy to translate into the higher redshift
bin, where (see Sect.\ref{sect:two-super} and
\ref{sect:color_seg_details}) we can explore only higher group
richnesses and higher galaxy stellar mass ranges.

For galaxies in the mass range $\log({\cal M}_{gal}/{\cal M_{\odot}}) > 10.56$
we do not observe any strong radial trend within groups in the mix of
red and blue galaxies ({\bf except possibly for poorer groups of our sample, 
where the $F_{\rm blue}$ value in the outermost region increases}) while we 
observe a clear difference with respect to \bluefrac for field galaxies.

For galaxies in the lower mass range $9.8 \leq \log({\cal M}_{gal}/{\cal M_{\odot}} <
10.56)$, we observe a significant radial dependence of the mix of red
and blue galaxies in poorer groups: red
galaxies are preferentially found in the group center. In contrast
the trend disappears for the richer group, suggesting that galaxies situated
in richer groups (\ie presumably located in more massive dark matter
halos) reach a redder color at earlier redshifts for each fixed galaxy
stellar mass.

This picture reflects what is found in the local Universe: most massive
galaxies do not show a significant color radial trend within groups,
while less massive galaxies are responsible for the progressive
blueing of the group member galaxies at intermediate and long
group-centric distances. Furthermore, these trends are stronger in
poorer environments \citep[see \eg][]{Bamford2009}.

Opposite to this differentiation in terms of a color segregation is 
the result shown in Fig. \ref{fig:radial_dist_cumul_rich}, implying
that a significant mass segregation is already set up in rich groups, 
whereas poor groups display a constant mix of galaxy stellar masses 
irrespective of the radial distance from group center. 

Do these differences provide the means to better understand how
the group environment influences the migration of galaxies from the
blue cloud to the red sequence, and to derive estimates for the
timescales involved in this process?

In other words, does the observed difference in
the radial color segregation between richer and poorer groups imply
the presence of a different efficiency in the two environments of the
mechanisms that cause the transition of a galaxy from the blue cloud
to the red sequence? Or could this difference just be caused by a
different time-scale for the evolution of structures of different mass
(different overdensity), such that richer groups formed earlier than
poorer ones, so that within them there simply has been more time for
environmental effects to act upon group member galaxies?

The observed difference in mass segregation trends provides an
important element for us to try and answer this question. It is well
known that mass segregation occurs when the exchange of energy among
group member galaxies has led most massive galaxies to set in the core
of the group while the lighter galaxies, moving at higher velocities,
preferentially reside in the outer regions.  The setting/absence of
mass segregation in a group therefore is a rough indication of the
time lapse since group formation, and because we do not dectect
any significant mass segregation in poor groups, these
systems are probably formed later than the richer ones.

This in turn may suggest that the absence of a color radial gradient in
richer groups at all masses we explored it is the result of the
longer time-scale of these systems, that is, of the longer time
available for the group environment to influence its member galaxies.

In contrast poorer groups have not yet been able to set mass
segregation, because they are possibly still in the process of forming and accreting
field galaxies. Their radial distribution in galaxy stellar masses
does therefore not show any strong segregation yet.  The radial color
trends we observe are somewhat reminescent of the still recent accretion
history of these groups and suggest that peripheral galaxies have been
accreted more recently than those located near the group center.

Interestingly, however, even the galaxies located at the outskirts
show redder colors than field galaxies of similar masses.  In
addition, the simple exercise we performed in
Sect.\ref{sect:near_far_field} shows that there is no continuous trend
in color segregation moving from the outskirts of groups toward the
nearest field: the physical scale on which the environment plays its
role coincides with the group physical scales, and the processes that
affect galaxy colors starts to operate as soon as galaxies enter the
group environment.  This result confirms that group environment
influences galaxy colors on short time-scales, in agreement with what
is suggested by the strong bimodality in color distribution itself:
any  quenching process that would last more than 1.5-2 Gyr would
erase the observed bimodality of galaxy colors by overpopulating green
valley \citep{Balogh2009,McCarthy2008,Font2008}.

Our analysis therefore implies that galaxy color transformation and
mass segregation originate from different physical processes, whose
time-scales, of a few Gyr, are only slightly different, so that
whenever mass segregation is observed, color segregation has already been
wiped out and viceversa.  Furthermore, it suggests that up to a
fixed galaxy stellar mass limit, galaxies have already been residing
for a longer time within richer groups compared to poorer groups, while for
each considered  group richness the galaxies of lower masses are presumably
those that have entered the group environment more recently (see
Fig.\ref{fig:F_blue_mass_rich}). The physical processes causing the
color trends observed in our data could be both starvation and/or
galaxy-galaxy collisions/interactions, because both operate on a similar
time-scale.  Starvation operates exclusively on the hot-gas
reservoir, \ie there are no indications that it results in 
structural transformation \citep{Weinmann2009}. 

On the other hand, galaxy-galaxy interactions can change morphology, boost
specific SFR (sSFR) and quench galaxies, and galaxy-pair fractions are
highly environmentally dependent, because denser environments show more
pairs \citep{Kampczyk2011}. An analysis including galaxy morphologies
and spectral features could help to understand which the most
likely process between the two is, because galaxy-galaxy collisions cause
morphological transformations while starvation produces the so called
'strangled' red-spiral population, and we plan to present it in a
forthcoming paper.

Our high-z results satisfactorily fit within the above picture presented for
low-z groups. \citet{McGee2009} show that at a fixed mass of
group/cluster considered, the accretion history of member galaxies is
remarkably similar and independent of redshift.

Given that the ${\cal M_{\mathrm{fudge}}}$ distributions of rich low-z
groups and high-z groups do not differ significantly, we would expect
a similar behavior in terms of both color and mass segregation. 

A KS test shows that the stellar mass distribution of galaxies more
massive than $\log({\cal M}_{gal}/{\cal M_{\odot}}) > 10.56$ does not
differ significantly for low-z rich groups and high-z groups and
in both cases they do not show any color radial trend.
Finally, both rich low-z groups and high-z
groups show evidence of mass segregation, confirming that the high-z
groups we observed in zCOSMOS do not deviate from the simple picture
we proposed.

\section{Conclusions}
\label{sect:Conclusions}

Taking advantage of the new \20K zCOSMOS spectroscopic data, its
excellent group catalog and the wide photometric coverage of the
COSMOS survey, we built two composite groups at intermediate (0.2
$\leq$ z $\leq$ 0.45) and high (0.45 $<$ z $\leq$ 0.8) redshifts. We
studied in detail how galaxy stellar masses and colors vary as a
function of the distance from the group center.  The analysis was
performed using mass-complete samples to separate the obvious
galaxy stellar mass/color dependencies. 

Our main results are:

(i) In the lowest redshift bin explored, the blue fraction of most massive
galaxies, \ie $\log({\cal M}_{gal}/{\cal M_{\odot}}) \geq 10.56$, does
not display strong group-centric dependence, despite displaying a
clear lower blue fraction in groups than in the field.
This result holds irrespective of group richness, 
except possibly for poorer groups of our sample, where
it is driven exclusively by the $F_{\rm blue}$ value in the outermost region, however.
In contrast, there is a radial dependence in the changing mix of red and blue
galaxies for galaxies of lower masses, 
\ie $9.8 \leq \log({\cal M}_{gal}/{\cal M_{\odot}}) < 10.56$,
with red galaxies being found preferentially in the group
center. This trend is stronger for poorer groups, while it 
disappears for richer groups.

(ii) In the highest redshift bin, where only higher galaxy stellar
masses and richer groups are available within the zCOSMOS survey, the
blue fraction of observed galaxies at $\log({\cal M}_{gal}/{\cal M_{\odot}})
\geq 10.56$ does not display strong group-centric dependence, although it
displays a lower blue fraction in groups than in the
field. 

(iii) The global ${F_{\rm blue}}$ for group galaxies shows a clear
dependence on group richness: rich group galaxies are redder than
poor group galaxies on average.

(iv) Mass segregation shows the opposite behavior with respect to
galaxy color trends: it is visible only in rich groups, while poorer
groups have a constant mix of galaxy stellar masses as a function of
group-centric distances.  Therefore the observed color trends cannot
be simply explained as caused by different stellar mass
distributions in different group regions.

(v) The physical length-scale on which the environment plays its role
coincides with the group physical scales.

The parallel absence(presence) of color segregation in rich(poor)
groups indicates that nurture effects are still in action in
poorer structures, whereas richer systems have already exhausted
their effects, so that all galaxies are uniformly red irrespective of
their position within the group (at least down to the galaxy stellar
masses we explored).  The corresponding presence(absence) in
rich(poor) groups of mass segregation suggests that richer systems
have been in place for long enough so that more massive galaxies
have sunk to the group center, something that has yet to happen for
the poorer groups, which still keep a memory of their more recent growth
history.

Both observations suggest a simple scenario where color and mass
segregation originates from different physical processes with similar
time-scales, so that whenever mass segregation is observed, color
segregation has been already wiped out and viceversa.

Lower mass galaxies in poorer groups are the witnesses of
environmental effects in action superimposed to secular galaxy
evolution: these galaxies still display gradually bluer colors moving
from group center to more external regions as a consequence of the
still recent accretion history of these groups. Starvation and
galaxy-galaxy interactions could both be the reason for a
mechanism that quenches star formation in groups at a faster rate than
in the field.

Future work will include a detailed analysis of galaxy morphologies
and composite spectra to investigate the scenario we presented in this 
paper in greater detail.

\begin{acknowledgements}
We are grateful to De Lucia, G., Girardi, M. and Andreon, S. for 
useful discussions and comments. V.P. acknowledges support from 
INAF-OAB PhD Grant.
\end{acknowledgements}

\bibliographystyle{aa}
\bibliography{biblio}

\begin{appendix} 
\section{The algorithm to add photo-zs}
\label{app:algorithm} 

In this appendix we present the details of the algorithm we adopted to
add photometric candidate group members to the spectroscopic ones,
thus recovering group members that were not observed spectroscopically
because of the incomplete sampling of \20K.

Given a group with N observed spectroscopic members, we define its
center on the sky: $ra_{gr}$ and $dec_{gr}$ center and its redshift
position, $z_{group}$, using the mean of the coordinates of its member
galaxies \citep[as in][]{Knobel2011}.

We then define $\mathrm{R_{gr}}$ as the minimum radius of the circle
centered on ($ra_{gr}$, $dec_{gr}$) containing all spectroscopic
confirmed members.

For each galaxy with apparent magnitude $I_{AB}\leq 22.5$ and photometric
redshift $z_{gal}$ we define
\begin{itemize}
 \item $\mathrm{R_{gal}}$, the projected radial distance of the galaxy
 to the center of the group on the sky;
 \item $\mathrm{x}=\mathrm{R_{gal}}/\mathrm{R_{gr}}$, the projected radial distance normalized to group size;
 \item $\Delta \mathrm{z}=|\mathrm{z_{gal}}-\mathrm{z_{group}}|$, the
 distance in redshift space to the position in redshift of the group;
 \item $\mathrm{y}=\Delta \mathrm{z}/\sigma_{\mathrm{zphot}}(I_{AB})$,
       the distance in redshift space normalized to
       $\sigma_{\mathrm{zphot}}(I_{AB})$, the photometric redshift
       accuracy computed at the galaxy magnitude $I_{AB}$.
\end{itemize}

To accept/reject a photo-z galaxy as group member we tested different
selection functions F(x,y), depending on the galaxy normalized radial
and redshift distances from group center and all satisfying the simple
empirical criterion that the accepted normalized distance in redshift 
space for a galaxy to be accepted as group member decreases at larger 
normalized radial distances.

Since one of our main goals is to recover the real richness of the
groups, we excluded selection functions that were too conservative or
too sharp in their radial dependence, producing a negligible increase
of interlopers to the group at the expense of a small gain in terms of
recovered real members.

After many trials we adopted a simple linear profile to associate
acceptable normalized distances in redshift space to radial distances
from group center on the sky. The formula we chose is y = 2-x, and
$x =1$ is the maximum distance to which the search was extended.  This
way the selection function linearly decreases the acceptable distance
in redshift space as we go away from the center up to
$\mathrm{R_{gr}}$.

For each group, we applied this selection function to all galaxies
with photometric redshifts and $\mathrm{I_{AB}}\le22.5$ located within
a cylinder of $\pm 2\times\sigma_{\mathrm{zphot}}$ depth and within a
region of inner radius $R_i$ and an outer radius $R_i + 0.2\times
R_{gr}$, increasing $R_i$ iteratively in steps of $0.2\times
\mathrm{R_{gr}}$ from zero to $\mathrm{R_{gr}}-0.2\times R_{gr}$.

We note that the 95\% of the missing real members are always confined
within $\pm2\times\sigma_{\mathrm{zphot}}(I_{AB})$, while the
interlopers are spread over the entire range
$\pm4\times\sigma_{\mathrm{zphot}}(I_{AB})$.  This is the reason why
we constrained the maximum redshift depth to
$\pm2\times\sigma_{\mathrm{zphot}}(I_{AB})$.  Furthermore, in the inner
regions the number of real missing members is always higher than that
of interlopers, while the ratio is reversed as we go far away from the center.

At the end of each run of the algorithm a new catalog of 20K+photo-z
member galaxies is produced.  If there was any multiple assignment for
a single photo-z we used a check function, $\mathrm{F_{check}}$, to
univocally assign it to a group.  The check function uses both the
phot-z information and the more reliable spatial information.  It is
defined as $\mathrm{F_{check}(x,y)}=\mathrm{x^2} \times \mathrm{y}$,
so that the photo-z is assigned to the group with the minimum
$\mathrm{F_{check}(x,y)}$.  Tests on simulations show that this check
function is able to recover the real membership for 74\% of photo-z
with multiple assignments to different groups.

Once the new catalog of group member galaxies is created, we
determine the new $ra_{gr}$ and $dec_{gr}$ centers, this time defined
using equation \ref{eq:VW_center}, while leaving $z_{group}$
unchanged, and the new $R_{gr}$ for each group using both
spectroscopic and newly added photometric members.

The whole algorithm then runs iteratively: the j$^{th}$ iteration uses
the center and radius of the (j-1)$^{th}$ iteration to search for
photo-z member galaxies.  We define the center-shift as the distance
between the j$^{th}$ center and the (j-1)$^{th}$ one. After two
iterations the center shift is less than 5\% of
$\mathrm{R^{j-1}_{gr}}$ for 90\% of the groups, meaning that the
centering for these groups has converged.  The third iteration is
enough for the centering to converge also for the remaining 10\% of
the groups.

Once applied to our data, our algorithm adds a total of 684 member
galaxies with photometric redshifts to the already existing 1437
spectroscopic groups member galaxies.

Reassuringly, the ratio of the number of spectroscopic redshift members
to the total number of members, \ie including member galaxies with
photometric redshift only, agrees well with the value $\sim
62\%$ of the median sampling rate within the central area, once we
take into account our completeness of $\sim 90\%$, as tested from
simulations (see Sect.\ref{sect:Mocks}).

If we had used the complete group catalog (spec+phot-z) as provided in
\citet{Knobel2011}, adding as photometric redshift members to each
group only the galaxies with an association probability $\geq 0.5$
(to easily reject multiple associations and select only the
most reliable phot-z members), we would have obtained a set of phot-z
member galaxies overlapping by $\sim 70\%$ with our photometric member
galaxies.  These common galaxies are assigned to the same group in
$\sim 95\%$ of the cases, while the group centers and richnesses are
not noticebly modified, differences in centering being in agreement
with our typical centering error (see Sect.\ref{sect:center}).

\end{appendix}

\end{document}